\documentclass{aa}

\usepackage{graphicx}
\usepackage{subcaption}
\usepackage{txfonts}
\usepackage{natbib}
\usepackage{amsmath}
\usepackage{multicol}
\usepackage{stfloats}
\usepackage{color}
\usepackage{yfonts}
\usepackage{xcolor}
\usepackage{ulem}

\usepackage[colorlinks=true, linkcolor=blue, citecolor=blue, filecolor=blue, urlcolor=blue]{hyperref}
\begin{document} 

   \title{Starspots on eclipsing giant stars I.}
   \subtitle{The sample and eclipse mapping examples}

\author{K.~Ol\'ah
          \inst{1,2}
        \and
          B.~Seli
          \inst{1,2,3}
        \and
          A.~Haris 
          \inst{4}
        \and
          S.~Rappaport
          \inst{5}  
        \and 
          M.~Tuomi
          \inst{4}
        \and R.~Gagliano
          \inst{6}
        \and T.~L.~Jacobs
          \inst{7}
        \and M.~H.~Kristiansen
          \inst{8}
        \and H.~M.~Schwengeler
          \inst{9}
        \and M.~Omohundro          
          \inst{9}
        \and I.~Terentev
          \inst{9} 
        \and A.~Vanderburg
          \inst{5}  
        \and B. Powell 
          \inst{10}
        \and V. Kostov 
          \inst{10,11} 
        \and Zs.~K\H{o}v\'ari
          \inst{1,2}        
}
   \institute{Konkoly Observatory, HUN-REN Research Centre for Astronomy and Earth Sciences, Konkoly Thege Mikl\'os \'ut 15-17., H-1121, Budapest, Hungary\\
              \email{olah@konkoly.hu}
        \and
            HUN-REN CSFK, MTA Centre of Excellence, Budapest, Konkoly Thege Mikl\'os út 15-17., H-1121, Hungary
        \and
            E\"otv\"os University, Department of Astronomy, Pf. 32, 1518 Budapest, Hungary 
        \and
            Department of Physics, P.O.~Box 64, 00014 University of Helsinki, Finland
        \and 
            Department of Physics, Kavli Institute for Astrophysics and Space Research, Massachusetts Institute of Technology, Cambridge, MA 02139, USA
        \and
            Amateur Astronomer, Glendale, AZ 85308, USA 
        \and
            Amateur Astronomer, Missouri City, TX 77459, USA
        \and
            Brorfelde Observatory, Observator Gyldenkernes Vej 7, DK-4340 Tølløse, Denmark
        \and
            Citizen Scientist, c/o Zooniverse, Department of Physics, University of Oxford, Denys Wilkinson Building, Keble Road, Oxford OX1 3RH, UK
        \and
            NASA Goddard Space Flight Center, 8800 Greenbelt Road, Greenbelt, MD 20771, USA
        \and
            SETI Institute, 189 Bernardo Avenue, Suite 200, Mountain View, CA 94043, USA
}            
        
   \date{Received January 15, 2025; accepted April 17, 2025}

 \abstract
{Spotted stars in eclipsing binary systems allow us to gather significant information about the stellar surface inhomogeneities that is otherwise impossible from only photometric data.
Starspots can be scanned using the eclipse (or transit) mapping technique, which takes advantage of the passage of a companion star (or planet) in front of a spotted giant star in a binary system.}
{Based on the characteristics of their ultra-precise space photometric light curves, we compile a list of eclipsing binaries whose primary component is a spotted subgiant or giant star, with the aim of applying the eclipse mapping technique to them.}
{Eclipsing binaries with giant primaries were selected from Transiting Exoplanet Survey Satellite (TESS) light curves by visual inspection. Spots showing up as bumps during eclipses are modeled with an eclipse mapping technique specialized for two stars, and the number of spots are found with the help of Bayes factors. The full light curves themselves were analyzed with time series spot modeling, and the results of the two approaches were compared.}
{We present a catalog of 29 eclipsing close binaries with active giant components and analyze TIC\,235934420, TIC\,271892852 and TIC\,326257590 from the Continuous Viewing Zones (CVZ) of TESS. Remarkable agreement is found between the starspot temperatures, sizes, and longitudes from the eclipse mapping results and the corresponding full light curve solutions. Spots are always present at the substellar points of the tidally locked binaries. Data from the TESS CVZ allow us to follow the changes of spot patterns on yearly timescales.}
{}

  \keywords{stars: activity --
            stars: late-type --
            stars: starspots --
            binaries: eclipsing
               }
   \maketitle

\section{Introduction}\label{intro}

Magnetic activity of giant stars has been well known and studied since the advent of stellar activity research \citep{1976ASSL...60..287H}, manifested in rotational modulation caused by starspots, showing chromospheric lines in emission, flares and activity cycles, that is, the same features as are observed in their main-sequence, or even pre-main-sequence counterparts. The binarity of stars has a crucial role in maintaining the magnetic activity: Most (but not all) of the well-studied active giant stars are members of binary systems \citep[see e.g.,][]{2017AN....338..903K}. However, uncovering the binary nature of giant stars in the absence of eclipses is not easy and requires radial velocity surveys.

The activity of giant stars, until recently, was mostly followed from rotational modulation and long-term variation (cycle-like behavior) of their light curves caused by starspots from ground-based, automated telescopes. The advantages of these data are their sometimes decades long time base and the different filters used for the simultaneous observations, giving extra information on the temperatures of spots. The shortcomings are lower accuracy and sporadic nature due to weather conditions and daytime interruptions of the data.  Flares, which would be a clear sign of magnetic activity, are hard to observe from the ground because of the high background luminosity of the giant star; thus, as to our knowledge, no systematic flare monitoring has been done at optical wavelengths. Nevertheless, flares have been observed in X-rays \citep[cf.][and references therein]{2024ApJ...961..130Z}, but this technique is not wide-spread, and no long-term time-series observations are dedicated to giant stars. 

Space photometry gave a new impetus to studying the activity of giant stars in a number of ways. The long-term photometric monitoring by space telescopes originally planned to detect planets, resulted as by-product in long, very high precision photometry of active giant stars (as of many other variable stars as well). Giant stars are in a state of rapid evolution, therefore they are much less numerous to observe. The four-year long continuous dataset of the {\it Kepler} satellite made it possible to find the true percentage of active giant stars in its field through finding rotational modulations and flares \citep{2021A&A...647A..62O}; and to study flare characteristics \citep{2022A&A...668A.101O}. For a recent review of observing stellar flares in the era of space photometry, see \citet{2024Univ...10..313V}.

The possibility to detect starspots with the help of eclipses in binaries with an active component was raised by \citet{1988Ap&SS.143....1B}, who gave proper equations to model a spot on a stellar surface eclipsed by the companion star; unfortunately, this idea has never been tested on real data. Next, \citet{1997MNRAS.287..556C} developed a different, surface integrating method for the problem, which was successfully used on photometry of XY\,UMa by \citet{1997MNRAS.287..567C} showing spots and their changes in time on the surface of the primary star in the system.

Searching for solar-like oscillations of giants in eclipsing binary systems \citet{2014ApJ...785....5G,2016ApJ...832..121G} found, that the giant primaries of the shortest orbital/rotational period systems show clear evidence of the magnetic activity during and outside the eclipses, and at the same time suppressed oscillations.

Observations of planetary transits soon resulted in finding occultations of starspots of active planet host stars. The first spot transit mapping was performed by \cite{2003ApJ...585L.147S} on HD~209458 using data from the Hubble Space Telescope (HST) 22 years ago. A few years later spots were discovered and modeled on the HST light curves of HD\,189733 by \citet{2007A&A...476.1347P}. It also became evident that characterizing the planets' parameters, the changing temperatures of the host stars due to starspots play important roles as was shown by \citet{2009A&A...505.1277C} and applied to real data by \citet{2009A&A...504..561W}. An extreme example of spot detection via planet transits was published by \citet{2021ApJ...907L...5A} who discovered no less than 198 spots on the surface of a K2V star from {\it Kepler} photometry. \citet{2025arXiv250218129H} found 105 spot eclipses by six planets after analyzing 3273 transits from 99 planets. 

An interesting application of planetary transits of starspots is presented by \citet{2010ApJ...723L.223W} and \citet{2011ApJ...733..127S}, showing that spot occultations by planets allow one to find the stellar obliquity, that is, the orientation of the stellar rotation axis relative to the planet's orbital axis. The method can also be used to find differential rotation and changing active latitudes on stellar surfaces, see \citet{2011ApJ...743...61S}.

To our knowledge, all previous attempts to characterize starspots using transits of planets or stellar companions, were done on main sequence stars. The reason behind this could be the long-lasting (several hours to days) and very shallow eclipses of the rare binaries with giant components, which are hard to observe from the ground.

However, by monitoring a large area of the sky by the Transiting Exoplanet Survey Satellite (TESS), it was possible to find such exciting and rare systems.  We checked the available TESS observations to find  eclipsing binaries which have active giant components, and part of their surfaces are scanned by the secondary stars during eclipses. 

The structure of the paper is as follows. In Sect.\,\ref{sample} we give a short description of the selected eclipsing binary stars with active giant primaries showing clear magnetic activity, and in Sect.\,\ref{sect_data} we provide a summary of the data collected for our study. The methods used are presented in the following Sect.\,\ref{sect_methods}. Our results are detailed in Sect.\,\ref{sect_results} and discussed in Sect.\,\ref{sect_disc}.

\section{Selection of the sample}\label{sample}

In the era of massive data sets from space photometry and automatized classification of objects from these data, it would seem that searching for a type of variable star by checking the light curves by eye is out of date. Still, the human eye-brain combination can recognize unusual patterns in the data that automated programs are often unable to flag. Members of the Visual Survey Group \citep[VSG,][]{2022PASP..134g4401K} have been surveying literally millions of light curves from the {\it Kepler}, {\it K2}, and TESS missions for more than than a decade now, and have found many astronomical treasures \citep[see, e.g.,][]{2022PASP..134g4401K}.  Most recently, these include the discovery of the shortest known outer period in a triple-star system of just 24.5 days \citep{2024ApJ...974...25K}.   

Indeed, our sample of eclipsing, active close binaries with giant primary stars might well not have been compiled without this visual surveying component.  Over the past five years, the VSG team has, among other targets, been specifically looking for binaries in the TESS light curves with the following properties: (i) large modulations due to spots; (ii) rotational modulations much larger than any ellipsoidal light variations, (iii) orbital periods longer than roughly five days, (iv) and with both eclipses being relatively shallow (e.g., $\lesssim$ few percent; the latter indicating a small unevolved star orbiting a giant). The surveying was done on (i) anonymous Full Frame Imaging (FFI) targets; (ii) Brian Powell's machine learning selected eclipsing binaries \citep[method discussed further in][] {2021AJ....161..162P}; and (iii) Chelsea Huang's Quick-Look Pipeline \citep[QLP,][]{2020RNAAS...4..204H}\footnote{\url{https://archive.stsci.edu/hlsp/qlp}}.  While conducting the survey, the VSG displays the light curves with {\sc LcTools} and {\sc LcViewer} software provided by \cite{2019arXiv191008034S}, which allows for an inspection of a typical light curve in just a matter of seconds.

\begin{figure}[thb]
    \includegraphics[width=\columnwidth]{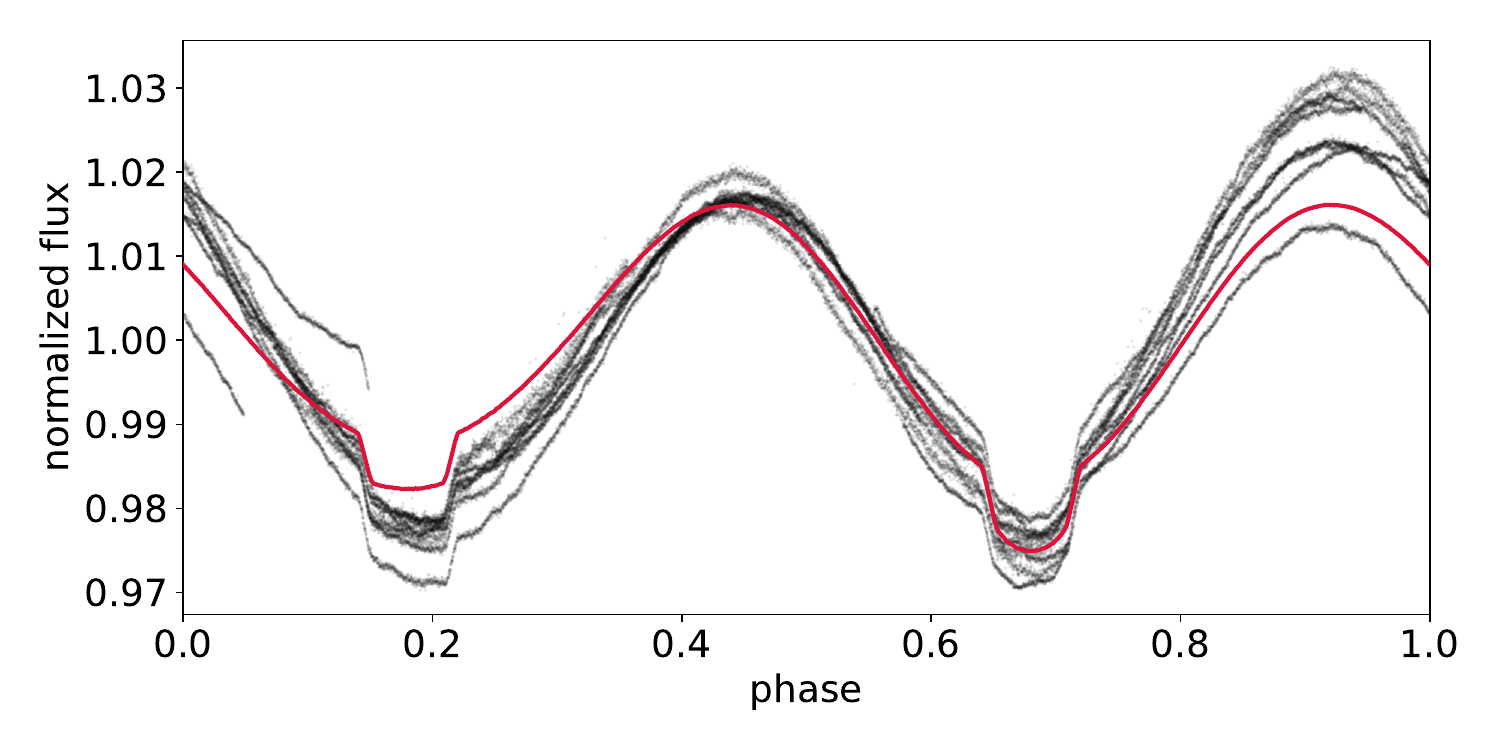}
      \caption{TESS light curve of V344\,Pup. The red line shows an approximate fit with PHOEBE without starspots, indicating ELV.}
      \label{v344Pup_TESS}
\end{figure}

Such a sample cannot be obtained from ground based data, since the amplitudes of the eclipses themselves are mostly below the observational accuracy under the Earth's atmosphere, the eclipses are long-lasting (several hours to days in different orbits), while the observational strategy is  usually one or a few datapoints per night with automated telescopes having also interruptions in the daylight. It could easily happen that a known active giant star in a binary system also shows eclipses which remained unnoticed until TESS observations. Such an example is V344\,Pup (Fig.\,\ref{v344Pup_TESS}) which is a known active giant star since 1987 \citep{1987SAAOC..11...21L} with a variable light curve both on longer and shorter timescales. An approximate PHOEBE \citep{2016ApJS..227...29P} eclipsing binary model fit to the TESS light curve and radial velocity data \citep{1987SAAOC..11....1B} shows that the light curve shape is dominated by ellipsoidal variations (ELVs), which, in this case, can explain most of the out-of-eclipse variation. Further, comprehensive information of the system is found in \citet{1998A&AS..131..321C}. 

Finally, our sample, together with V344\,Pup, consists of 29 eclipsing binaries with active giant components, as listed in Table\,\ref{basic_data} together with their basic parameters from version 8.2 of the TESS Input Catalog \citep[TIC,][]{2019AJ....158..138S}.

\begin{table}[thb]
\small
\centering
\caption{Basic data of the sample from TICv8.2}
\label{basic_data}
\begin{tabular}{r r r r r}
\hline
\hline\noalign{\smallskip}
TIC &     TESS & Distance & $T_\mathrm{eff}$ &  Radius \\ 
 & mag. & [pc]  & [K] & [$R_{\odot}$]\\
\hline                
\noalign{\smallskip}                                 
28124546  &   9.20  & 657 & 4888 & 9.6 \\
52584838  &   11.52 & 525 & 4433 & 3.5 \\
67516658  &   10.65 & 455 & 4585 & 3.9 \\
84194965  &   12.79 & 1132 & 4520 & 4.2 \\
92771205  &   11.97 & 527 & 4430 & 2.8 \\
103573024 &   8.70    & 445  & 4753 & 8.9 \\
116745617 &   10.68 & 1048 & 4996 & 8.6 \\
130472153 &   10.02 & 958 & 4845 & 10.4 \\
150392949 &   13.17 & 1663 & 5031 & 3.7 \\
153859909 &   10.47 & 1328 & 4703 & 11.6 \\
161573242 &   12.23  & 988 & 4697 & 3.8 \\
166974938 &   11.31  & 971 & 5233 & 4.7 \\
229950558 &   12.27 & 773  & 4477 & 3.4 \\
231490250 &   11.43 & 1221 & 4875 & 6.8 \\
235934420 &   12.63 & 862 & 4482 & 3.3 \\
255926044 &   12.95 & 1011 & 4458 & 4.1 \\
264195084 &   11.78 & 661 & 4534 & 3.6 \\
271892852 &   12.45 & 651 & 4873 & 2.4 \\
284889425 &   10.23 & 993 & 4453 & 11.6 \\
290552720 &   11.19 & 971 & 5000 & 5.7 \\
300301114 &   11.28 &         &      &         \\
304131827 &   12.28  & 1369 & 4769 & 5.1 \\
326257590 &   10.36 & 397 & 4526 & 4.2 \\
355367823 &   10.40 & 436 & 4525 & 4.2 \\
356837516 &   11.02 & 774 & 4669 & 5.9 \\
396059992 &   8.86  & 199 & 5103 & 3.1 \\
441804674 &   9.03  & 318 & 4684 & 5.4 \\
470244846 &   12.58 & 2090  & 4425 & 9.0 \\
\hline                                               
123153249 &   5.99  & 119 & 4815 & 7.8 \\
\hline
\end{tabular}
\end{table}

\begin{figure}[thb]
    \includegraphics[width=\columnwidth]{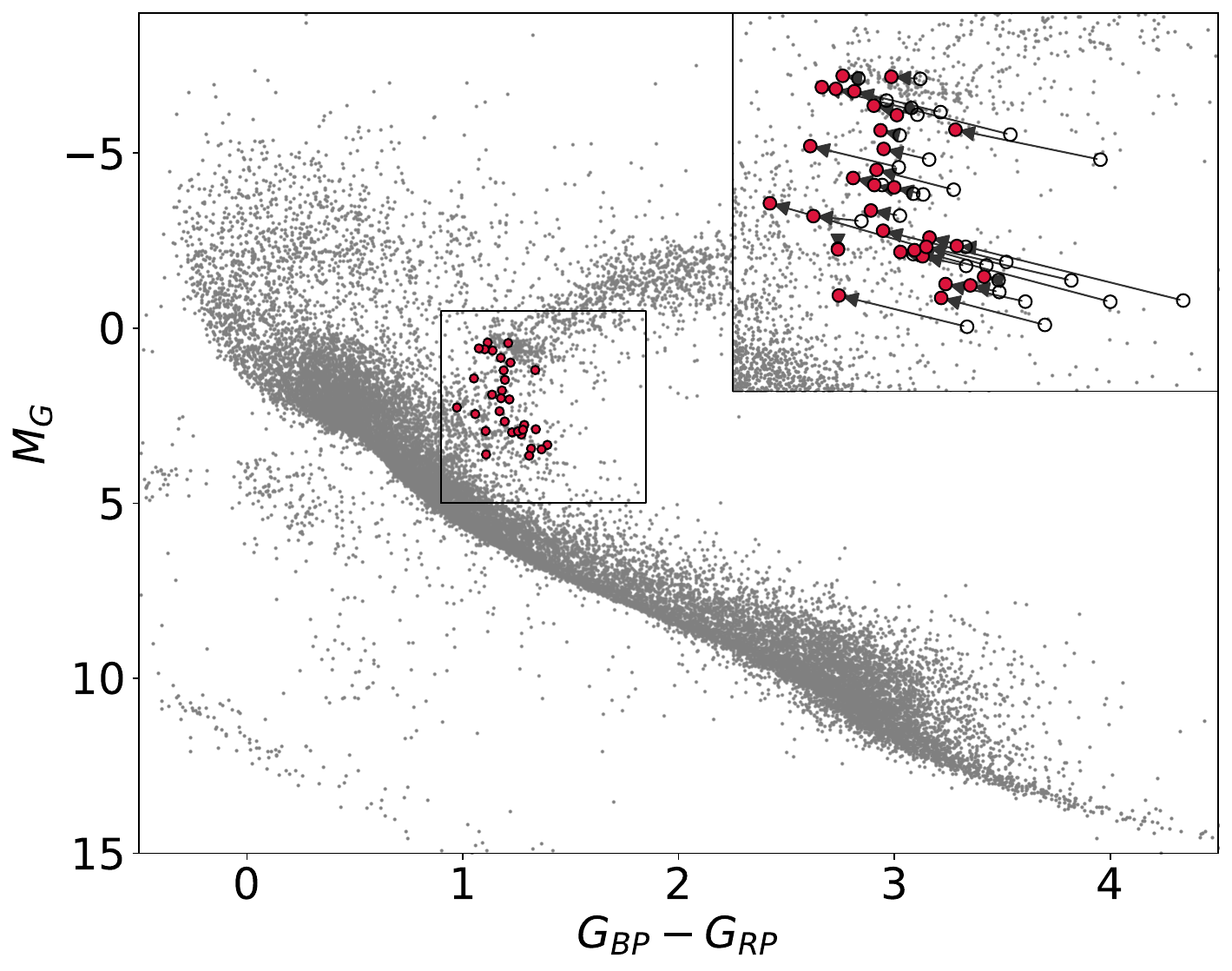}
      \caption{Sample binaries on the Gaia DR3 color-magnitude diagram. Red circles show the values after the extinction and reddening correction, the empty circles in the inset are the original uncorrected values. The gray points show a representative sample of the solar neighborhood for reference.}
      \label{HRD}
\end{figure}

Figure\,\ref{HRD} shows the position of the sample on the Gaia DR3 color-magnitude diagram. To retrieve the magnitudes corrected for galactic extinction and reddening, we used the \texttt{seismolab} package\footnote{\url{https://seismolab.readthedocs.io/en/latest/index.html}} with distances from \citet{2021AJ....161..147B} and the default `Combined19' dust map.

From the final set of binaries that fulfilled the selection criteria it turned out that most of them (with only one exception) show unambiguous signatures of spot eclipses by the secondary stars. In a small series of papers we will investigate what we can learn about starspot activity in these close binaries with giant components. In this first paper we study observations of three binaries, TIC\,235934420, TIC\,271892852 and TIC\,326257590, of the four from our sample that are located in the Continuous Viewing Zones (CVZ) of TESS. As an example, Fig.\,\ref{fig:example_lc} depicts a series of light curves of TIC\,326257590 with one smooth transit eclipse when the giant star is in front. Half an orbit later the secondary star scans a part of the surface of the giant showing bumps when crossing spots which apparently slowly vary in time. Flares, as other activity signatures are also seen on the light curves. We present time-series spot modeling parallel with a novel approach of eclipse mapping of spots by the secondary stars.

\section{Data}\label{sect_data}

\subsection{TESS}\label{TESSdata}

The basis of the present work is the database of TESS \citep{2014SPIE.9143E..20R}. We used the \texttt{lightkurve} python package \citep{lightkurve2018} to download TESS data products of our sample. We downloaded the shortest available cadence light curves for each TESS sector created by the Science Processing Operations Center \citep[SPOC,][]{10.1117/12.2233418} pipeline. If SPOC observations were not available for a sector, we took the observations reduced by QLP \citep{2020RNAAS...4..204H}. We used observations processed with the Pre-search Data Conditioning Simple Aperture Photometry (\mbox{PDCSAP}) method. The cadences of the observations were 200-s, 600s and 1800s for TIC\,235934420, 200-s and 600-s for TIC\,271892852, and 120-s, 200s and 1800s for TIC\,326257590. All cadences allowed us to carry out eclipse mappings when there was no problem with the data. Specifically, the cadences of the data used for the plotted results are given in the respective section of each star.

\begin{figure}[thb]
    \includegraphics[width=\columnwidth]{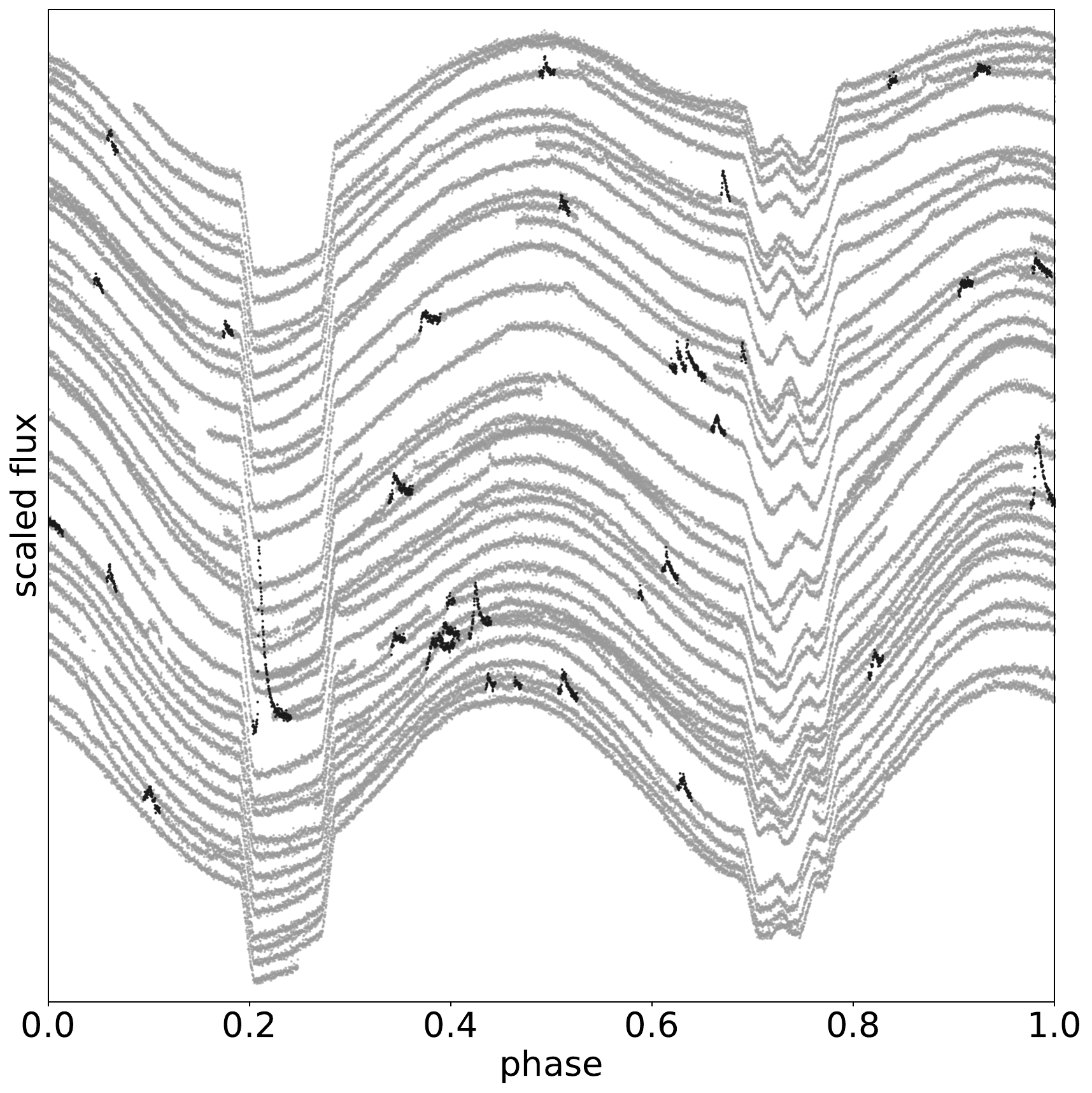}
      \caption {Observed TESS light curve of TIC\,326257590. The light curve is folded with the orbital period of 6.0518\,days and shifted in time from top to bottom, showing spot crossings by the secondary star during eclipses. Flare events are marked.
      }
      \label{fig:example_lc}
\end{figure}

\subsection{ZTF}\label{ZTFdata}
The fully automatic Zwicky Transient Facility (ZTF) is a survey systematically exploring the optical transients in the northern hemisphere \citep{2019PASP..131a8003M}. Its wide-field camera on the 48-inch Schmidt telescope at Palomar Observatory provided high quality multicolor ($g$, $r$, $i$) observations for TIC\,235934420.

\subsection{ASAS SN}\label{ASASdata}
We made use of the public photometry of the All-Sky Automated Survey for Supernovae (ASAS\,SN). It consists of 24 telescopes which automatically survey the entire visible sky every night down to about 18 magnitude \citep{2014AAS...22323603S}.

\section{Methods}\label{sect_methods}

\subsection{Transit light curves}\label{BLS}

To obtain the orbital period of each system, we first looked for a tentative orbital period $P_{\mathrm{guess}}$, which yielded overlapping transits in the phase folded light curves of the binaries.
We divided the TESS observations into sections of length $P_{\mathrm{guess}}/4$ and normalized each section with a cubic spline fit.

We then looked for the orbital period $P_{\mathrm{orb}}$ in the normalized light curves with the Box Fitting Least squares algorithm \citep[BLS,][]{kovacs2002} within the range ($P_{\mathrm{guess}}$$-1/2$\,d, $P_{\mathrm{guess}}$$+1/2$\,d). We accepted the value that maximized the likelihood function of the fit as the orbital period of the system. We took the distribution of the transit mid-times from the MCMC sampling of the transit light curves in BJDs. We then sampled a linear regression on these mid-transit times, with MCMC to obtain the error estimates for the orbital periods $P_\mathrm{orb}$ and transit epochs $T\mathrm{_0}$.

Some stars were observed by TESS with different cadences in different sectors. Since sectors with a shorter cadence have more datapoints, the BLS analysis would favor those sectors with a denser sampling and may yield imprecise period estimates. In order to avoid this scenario, we rebinned every TESS sector of each star down to the longest cadence 1800-s observations. 

While the BLS analysis yields high precision orbital period estimates, it only crudely estimates the mid-time of the first transit $T_0$ and the length of the transit $T_{14}$. To estimate these two parameters more precisely, we visually inspected the first available TESS sector of each binary in our sample, and fine-tuned the parameters manually. We then calculated the mid-times for all primary and secondary transits between the start and end time of the observations. We refer to the transit of the smaller component in front of the bigger component as the primary transit.
We assumed that the primary and secondary transits are separated by $P/2$, which at high time resolution, corresponds to circular orbits.

We extracted all observations within $T_{14}$ of all transit mid-times (both primary and secondary). We also assumed all orbits to be circular. This assumption has minimal effects on the results, because transit data provides rather weak eccentricity constraints \citep{2015ApJ...808..126V}. Similarly to the procedure described by \citet{2025arXiv250218129H}, we shifted and divided each transit by its difference to the mean flux, and normalized the out-of-transit baseline of the light curves by a third order polynomial.

\subsection{Light curve model}

\begin{figure}[thb]
    \includegraphics[width=0.5\columnwidth]{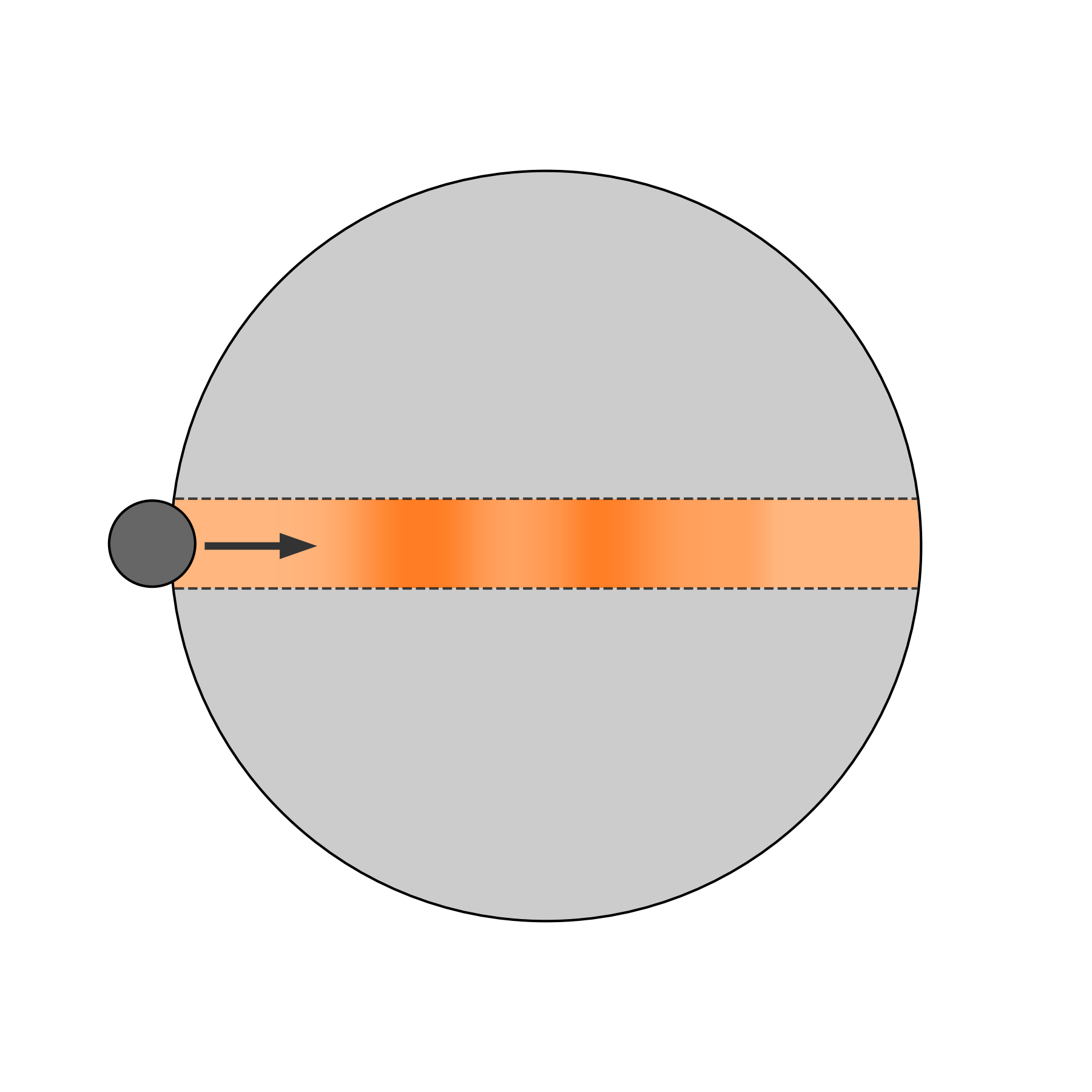}\includegraphics[width=0.5\columnwidth]{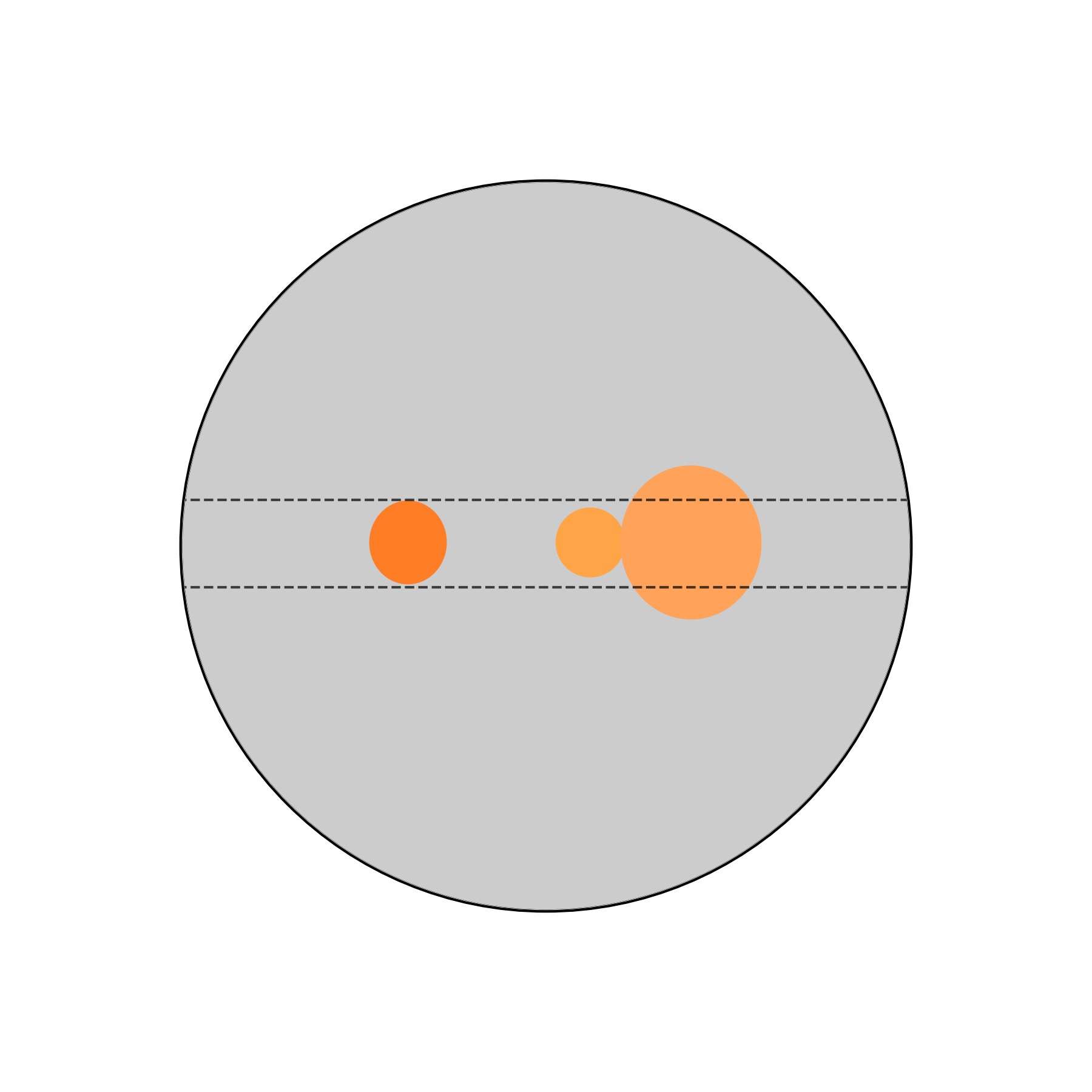}
      \caption{Spot eclipses on the surface of a giant star by its secondary. The horizontal belt shows the scanned latitudes. Left: the path of the secondary star as it moves through the spots shown by the color coded pattern of the photometry during eclipses, right: the resulting three spots after the solution. The coloring of the features in both panels reflect the different contrasts of the spots to the stellar photospheres.}
      \label{fig:example_disk}
\end{figure}

Let us take an eclipsing binary system on concentric, circular orbits, assuming spherically symmetric components with no mass-transfer between them. The flux ratio of the stars at some observed wavelength is $k_{f} (\lambda) = f_{p}/f_{s}$, their radius ratio is $k_r = r_p/r_s$. 

To describe the change in flux for the binary system, we make slight modifications to the planetary transit equations of \citet{mandel2002}. We refer to the inherent/uneclipsed flux of the eclipsed component at time $t_i$ as $F(t_i)$, as defined by \citet{mandel2002}. The normalized flux of the binary system can be described as
\begin{equation}
    f(t_i) = \frac{1}{f_p + f_s} F_{\mathrm{phase}}(t_i)\,,
\end{equation}
where the phase dependent flux of the system is
\begin{equation}
F_{\mathrm{phase}}(t_i) = \left\{ \begin{array}{cl}
f_p \, F(t_i) + f_s\,, &  \mathrm{primary \ transit} \\
f_s \, F(t_i) + f_p\,, & \mathrm{secondary \ transit} \\
f_s + f_p\,, &  \mathrm{out-of-transit}\,.
\end{array} \right.
\end{equation}

In this formulation, the binary system is defined by the flux ratio of the two bodies observed at the central wavelength of TESS $k_f = f_p/f_s$, the radius ratio of the two bodies $k_r$, the inclination of the shared orbit, the distance of the two bodies as a multiple of the radius of the primary component $d/r_1$, the orbital period of the system, and two quadratic limb darkening parameters ($u_1, u_2$) per star. We assumed that the observations are contaminated by some unknown noise, following a Gaussian distribution $\epsilon_i \sim N(0,\sigma_{\mathrm{G}}^{2})$.
We used independent mid-transit time $t_c$ parameters for all transits, to account for possible transit timing variations. The binary transit model had thus $k_{\mathrm{binary}}= 9 + n_{\mathrm{transit}}$ parameters, where $n_{\mathrm{transit}}$ is the number of transits.

To model spot eclipses we follow the practice of \citet{tuomi2024} and \citet{2025arXiv250218129H}. The signature of a spot eclipse in the transit light curve $g_{\mathrm{spot}}$ can be approximated by a `plateau model', in which the observed brightness increases and decreases exponentially, interrupted by a constant brightness phase:

    \begin{equation}
    \label{eq:spot_eclipse}
        g_{\mathrm{spot}}(t_i) =
        \left\{\begin{matrix}
        A \exp \left [ -\frac{(t_i-t_{\mathrm{in}})^2}{2 \sigma_{\mathrm{spot}}^2} \right ]\,, & t_i < t_{\mathrm{in}}\\ 
        A\,, & t_{\mathrm{in}} < t_i < t_{\mathrm{out}} \\ 
        A \exp \left [ -\frac{(t_i-t_{\mathrm{out}})^2}{2 \sigma_{\mathrm{spot}}^2} \right ]\,, & t_{\mathrm{out}} < t_i
        \end{matrix}\right.
    \end{equation}
where $A$ is the amplitude of the spot-induced flux change, $t_{\mathrm{in}}$ and $t_{\mathrm{out}}$ are the starting and end times of the spot transit, and $\sigma_{\mathrm{spot}}$ is the timescale of the flux change. When the stars do not transit each other, $g_{\rm spot} (t_i) = 0$. A model with $n_{\rm spot}$ spots is
\begin{equation}
    m_i = f(t_i) + \sum_{j=1}^{n_{\rm spot}}  g_{\mathrm{spot},j}(t_i) + \epsilon_i\,.
\end{equation}
To minimize computational cost, we fixed the orbital period to the value obtained from the BLS analysis. Additionally, we assumed all orbits to be circular, as the primary and secondary transits of the objects occurred with a periodicity very close to $P_{\rm orb}/2$.
Figure\,\ref{fig:example_disk} visualizes the spot eclipses and results based on real data.

We note that by using this formulation of the spot transit signal, we have not assumed any specific spot shape yet (e.g., circular). Using a single eclipse, we can only observe a chord on the stellar surface, sometimes passing through starspots (see Fig.\,\ref{fig:example_disk}). In the following sections, for illustration purposes, we will assume that the measured signal comes from \textit{central} occultation of \textit{circular} spots. Thus, the spot sizes should be treated as lower limits.

There are different approaches in the literature to model spot eclipses, most of them assume circular spots \citep[e.g.,][]{2008ApJ...683L.179S, 2009A&A...504..561W, 2015csss...18..399D, 2023ApJ...956..141A}.
Another method was presented by \citet{2019AJ....157...64L} called \texttt{starry}, which uses spherical harmonic decomposition to reconstruct the stellar surface, utilizing the full light curve to constrain the spherical harmonic coefficients.

\subsection{Posterior sampling}

To sample the probability distribution of the parameters that describe the light curve of the binary system and the possible spot eclipses, and to assess the significance of spot eclipse signals,
we use the Adaptive Metropolis (AM) algorithm of \citet{haario2001}. The AM algorithm is a Markov Chain Monte Carlo (MCMC) method, which is an updated version of the Metropolis$-$Hastings algorithm \citep[MH,][]{metropolis1953, hastings1970}, in which, after each iteration, the proposal distribution is changed to the covariance matrix of the vectors generated so far. This enables a much faster convergence around the posterior distribution than the classic MH algorithm.

Sampling was conducted independently for each TESS sector. The light curves around primary and secondary minima within a single sector were sampled together. We used one Gaussian noise term $\sigma_{\mathrm{G}}$ per TESS sector. For each sector, models with $0$ to $4$ spots were sampled.  Similarly to \citet{2025arXiv250218129H}, we used uniform, uninformative priors for each parameter $\theta$ such that we set them to unity in some interval $\theta \in [a,b]$ and zero outside that interval. We drew initial values for sampling randomly from the proposal distribution, except for inclination, which was chosen to be high enough to ensure that the two components of the binary eclipse each other.

When setting the range of priors for the spot parameters, we require that the amplitude of the spot eclipse is smaller than the transit depth. This minimizes the risk of interpreting other astrophysical phenomena as spot features. The maximum amplitude phase of the spot eclipse signal (between $t_{\rm in}$ and $t_{\rm out}$) is constrained to less than half of the transit duration. To avoid the overlap of dark regions, we required that the maximum amplitude phases of different spot signals are detached. Table\,\ref{tab:params_sampling} presents the intervals of the priors.

\begin{table}[t]
\caption{Limits for uniform priors $a,b$, used for the posterior sampling.}
\label{tab:params_sampling}
\centering
\begin{tabular}{@{}lcc@{}}
\hline
\hline
Parameter & $a$ & $b$ \\ \hline
$k_f$ & 1 & 100 \\
$k_r$ & 0.1 & 15 \\
$i~(^\circ)$ & 75 & 90 \\
$d/r$ & 1 & 10 \\
$u_{*,1}$ & $\mathrm{max}(0, -2 u_{*,2})$ & $1 - u_{*,2}$ \\
$u_{*,2}$ &$-1/2 \, u_{*,1}$ & $1-u_{*,1}$ \\
$\sigma_{\mathrm{G}}$ & $0$ & $1$ \\
$t_c$ & $-T_{14}/10$ & $T_{14}/10$ \\
$A$ & $0$ & $\delta$ \\
$t_{\mathrm{in}}$ & $t_c-T_{14}/2$ & $t_{\mathrm{out}}-120~\mathrm{s}$ \\
$t_{\mathrm{out}}$ & $t_{\mathrm{in}}+120~\mathrm{s}$ & $\mathrm{min}(t_{\mathrm{in}}+T_{\rm in}/2$, $t_c+T_{14}/2)$ \\
$\sigma_{\mathrm{spot}}$ & $0$ & $T_{14}/10$ \\ \hline
\end{tabular}
\end{table}

The number of sampled parameters for each TESS sector was
\begin{equation}
k = 9 + n_{\mathrm{transit}} + 4\,n_{\mathrm{spot}}\,n_{\mathrm{primary}},
\end{equation}
where $n_{\mathrm{spot}}$ is the number of spots assumed by the model, and $n_{\mathrm{primary}}$ is the number of primary transits. We sampled models with spot numbers $n_{\rm spot} \in [0,4]$, to ensure that all detectable spots are identified.
To verify that the posterior distribution has been correctly identified, we sampled the posterior with 2--4 chains simultaneously. A low number of chains was sufficient, as spot eclipses usually leave unambiguous signals in the time-series with well constrained, unique solutions in the parameter space. We required that the Gelman$-$Rubin statistic of each parameter was below 1.05 to claim that there was no evidence for the non-convergence of the chains and that the identified solutions were unique.

\subsection{Model comparison}

Some light curves have longer cadences or exhibit higher noise levels, making the spot signals more ambiguous and preventing a definitive determination of the number of spots through visual inspection alone. Therefore, an objective set of criteria is necessary to determine how many spots were eclipsed during the transit. This approach was also used by \citet{tuomi2024} and \citet{2025arXiv250218129H} to determine the number of eclipsed spots in the transit light curves of exoplanets.

To evaluate the significance of different models, we compute the Bayesian information criterion (BIC) for a model with $n_{\rm spot}$ spots, applied separately to each primary minima. The BIC values are then used to estimate the probability of each model, $P(m)$. The BIC value is calculated as
\begin{equation}
\mathrm{BIC} = -2 \log P (m) = - 2 \log l_{\mathrm{max}} +k \log N
\end{equation}
where $l_{\mathrm{max}}$ is the maximum likelihood of the model, found during the sampling of the posterior, $k$ is the number of free parameters, and $N$ is the number of datapoints. We note that this approximation is only valid if the number of observations is high and if the probability density function is in the vicinity of the maximum likelihood estimate. 

The Bayes factor in favor of model $\mathcal{M}_i$ against model $\mathcal{M}_j$ is defined as \citep{kass1995}
\begin{equation}
B_{i,j}(m) = \frac{P(m|\mathcal{M}_i)}{P(m|\mathcal{M}_j)}\,.
\end{equation}

We adopt the scale of \citet{jeffreys1961} and \citet{kass1995} for assessing model probabilities, based on their Bayes factors. Following this scale, the Bayes factor of a model with $n+1$ spots against a model with $n$ has to be at least $B_{n+1,n} > 150$ for a decisive evidence in favor of $n+1$ spots.

\subsection{Inferred spot properties}

Modeling the geometrical properties of the eclipsed regions for eclipsing binaries bears a close resemblance to the same modeling from exoplanetary transits. For this reason, we adopted the equations from \citet{2008ApJ...683L.179S} to calculate the latitude, longitude, and radius for the eclipsed regions.

We define the contrast of the eclipsed region as the ratio of the amplitude of the spot signal to the depth of the unspotted transit light curve at the mid-time of the spot occultation event, expressed as $c = A/\delta(t_{\rm mid})$.
In the figures later, we plot the whole area of the resulting spot including the `plateau' and the decrease to the base level (Eq.\,\ref{eq:spot_eclipse}). 

The resulting spot longitudes are not exactly the same near the eclipses from eclipse mapping and from time-series spot modeling (see Sec.\,\ref{ts_spotmodel}). In the course of eclipse mapping, due to the optimization of all free parameters together, the time of the eclipse minimum is slightly changing. Comparing those minima to the calculated ones the difference is only a few degrees at most, which is negligible when interpreting the resulting spot configurations.

Inherent uncertainties come from the durations of the eclipses, from the brightness of the objects through signal-to-noise (S/N), and from the cadences of the observations. These properties together determine the resolutions of the maps. 

The detection statistics of spot eclipse signals are strongly dependent on which section of the active region is transited by the eclipsing body. Eclipses in which the eclipsing body only grazes the spot without transiting its central latitude are less likely to be confirmed as a spot by our detection criteria. More discussion of spot detection limitations is found in Fig. 1. and Appendix C in \cite{2025arXiv250218129H}. 

\begin{figure}[h!!!]
   \includegraphics[width=0.5\columnwidth]{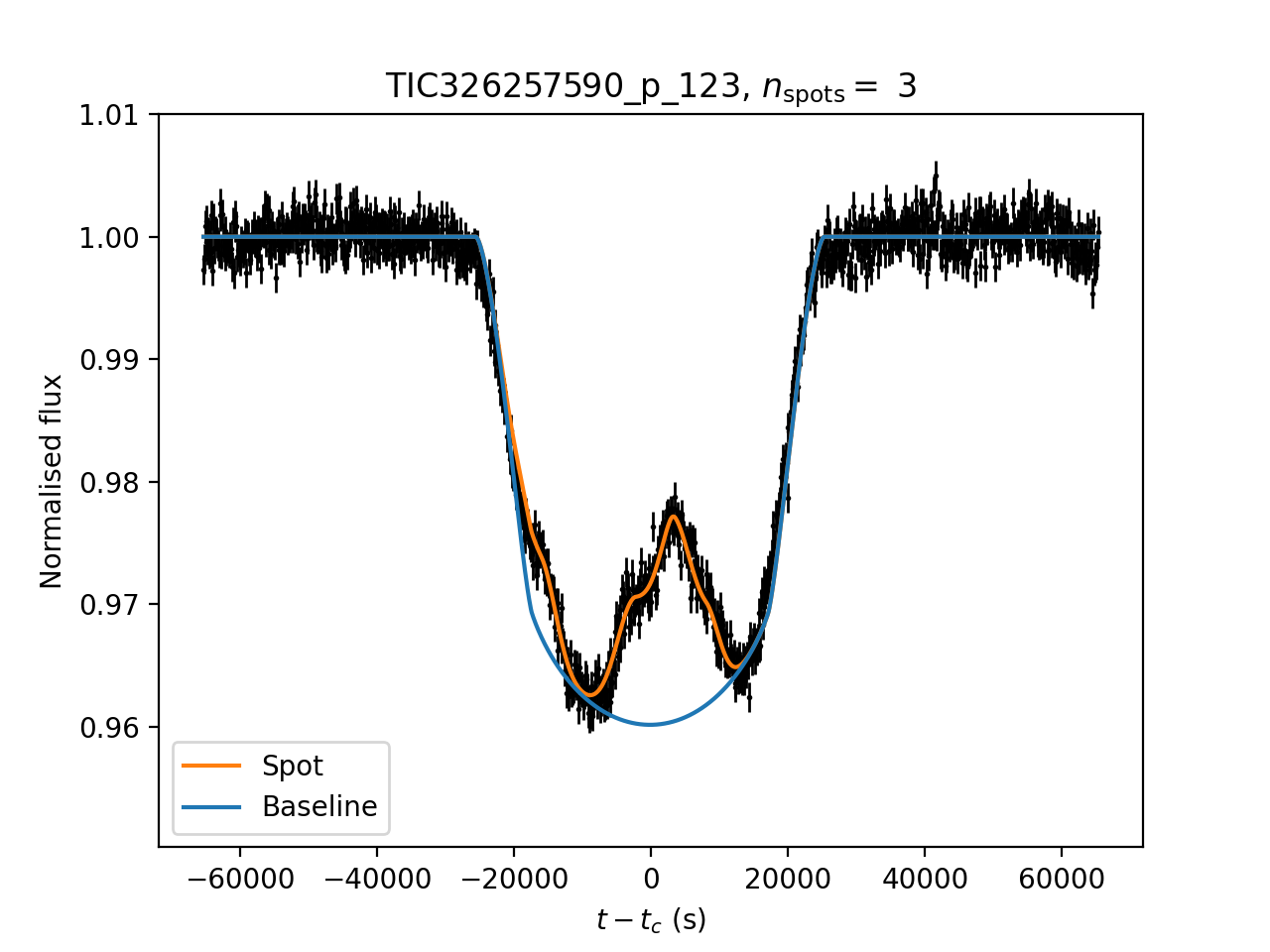}\includegraphics[width=0.5\columnwidth]{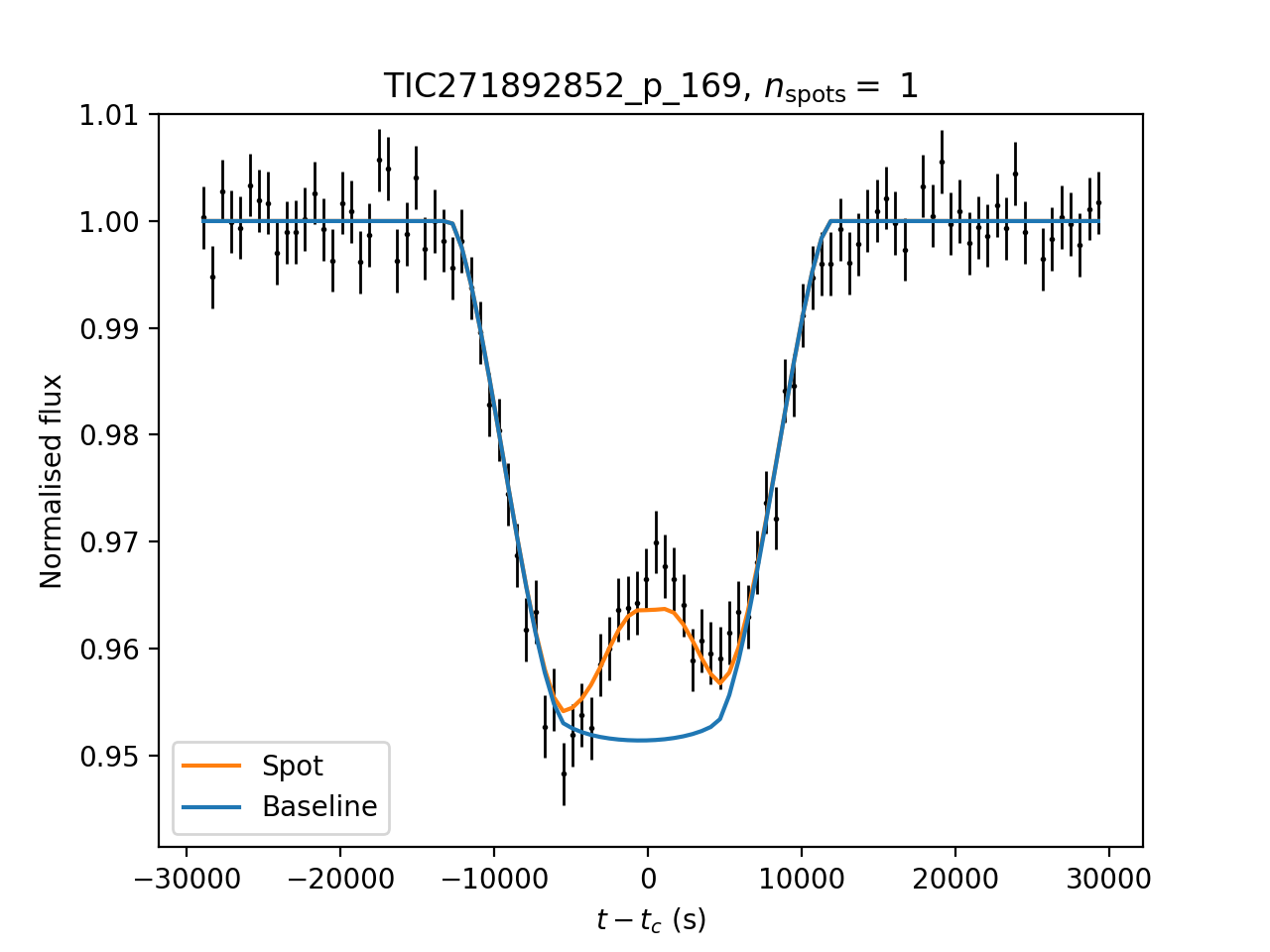}
      \caption{Working figures of eclipse mapping on two different targets. Plotted are the normalized fluxes as a function of time expressed in seconds counted from the mid-eclipse. Left: TIC\,326257590, right: TIC\,271892852. See the text for explanation.} 
      \label{fitexample}
\end{figure}

In the following example there is an approximately 2 magnitudes difference between TIC\,326257590 and TIC\,271892852 (Table\,\ref{basic_data}), observed with 120-s and 600-s cadence, respectively. The fitted minima are displayed on Fig.\,\ref{fitexample}. The durations of the minima are different. The scanned latitudes where the secondary passes the primary is around the equator for TIC\,326257590 and at about +40$^{\circ}$ for TIC\,271892852 (Table\,\ref{3stars}), which affect the shapes of the eclipses. Looking at Fig.\,\ref{fitexample} one can see the difference between the details of the fit results due to the factors detailed above, on the two, similar looking spot signals. 

\subsection{Time-series spot modeling of the light curves}\label{ts_spotmodel}

The light variations on rotational and longer timescales of the active component of the binaries were modeled using our own software based on the analytical equations applied to spot modeling assuming circular spots, by \citet{1977Ap&SS..48..207B}, in time-series mode. Such a modeling is an ill-posed problem, since we try to construct two-dimensional maps from one-dimensional (photometric) data. We have very limited latitude information especially due to high inclinations (close to 90$^{\circ}$) of the spotted stars in the detached eclipsing binaries, therefore constant latitudes were assumed around the scanned latitudes of the giants by the secondary stars. Longitudes could be found with fairly high precision, the sizes of the spots depend on the supposed spot temperatures and also on the positions on the stellar surface. Accuracy of the light curves also plays a role in the stability of the resultant spot parameters; for an early investigation of the problem see \citet{1997A&A...323..801K}. In Appendix\,\ref{spotlatitudes} we present experiments about recovering some (limited) information on spot latitudes.

The light loss by spots should obviously be measured from the unspotted brightness of the stars, which is usually unknown. Instead, we use the observed maximum brightness, so the inferred and modeled spots account for the light curves themselves, assuming that the remainder of the dimming is caused by evenly distributed and/or polar spots. 

The single color TESS data have no information on the temperature of the starspots. The resulting spot areas are most probably activity complexes, where spot umbrae, penumbrae and faculae are present together, and only their summed up effect is observed as a cool region. At present it is not possible to separate these regions in stars, which are also subject to changes in their respective areas with time. Thus, the term `spot temperature' means a gross average temperature value of a region we call `starspot'.  

The problem of the spot contrasts from TESS data alone lies in the lack of knowing the unspotted brightness of the star, and is described by \citet[][see their Fig.\,11 for a clear explanation]{2021AJ....162..123L} saying that `...total spot coverage of a star cannot be uniquely inferred from single-band photometry'.

From multicolor data spot temperatures, that is, average temperatures of the active regions, can be derived. In most cases, for the one-color TESS data, we had to suppose constant, typical spot temperatures based on the eclipse mapping results. For one binary in the sample we have contemporaneous TESS and two-color ZTF data, therefore we had the possibility to compare the resulting spot temperature with that from the eclipse mapping, see Appendix\,\ref{ZTF_two-color}. 

For the time-series spot modeling a linear limb-darkening coefficient $u= 0.66$ was used from \cite{2017A&A...600A..30C} for the TESS bandpass for all stars. This is a compromise, since we do not have precise stellar temperatures, and no reliable information for log$g$ and metallicity of the objects, moreover, the uncertainties are of the order of 0.1-0.3. The value used is similar to the quadratic part of the eclipse mapping results within the uncertainties.

All eclipses and flares were removed from the data before spot modeling of the light curves. In modeling the high S/N TESS data we assumed 3$-$4 spots, which means 6$-$8 free parameters of longitudes and radii depending on the quality of the light curves, while the latitudes were kept constant. In the case of the ZTF data three spots were suitable to find acceptable results with seven free parameters, that is, the longitudes and radii of the three spots, and the spot temperature; see Appendices\,\ref{spotnumber} and \ref{ZTF_two-color} for more.

\section{Results}\label{sect_results}

\begin{table*}[thb]
\small
\renewcommand{\arraystretch}{1.5}
\setlength{\tabcolsep}{4pt} 
\centering
\caption{Data of the discussed giants and binary orbits based on TESS data analysis}
\label{3stars}
\begin{tabular}{ccccccccc }
\hline\hline                
\noalign{\smallskip}
TIC & $P_\mathrm{rot}$ & $P_\mathrm{orb}$ and $T_\mathrm{0}$ & $k_f$ & $R_1/R_2$ & $i$          & $a/R_1$ & scanned          & scanned \vspace{-5pt}\\ 
No. & [d]             & [d]               &       &           & [$^{\circ}$] &         & latitude range        &   longitude range \\
\hline
235934420 & 7.2379$\pm$0.0115 & $7.23710_{-0.00014}^{+0.00014}$ & 29.30$_{-0.83}^{+1.28}$ &  5.36$_{-0.15}^{+0.12}$ & 88.68$_{-1.12}^{+0.90}$ & 5.24$_{-0.08}^{+0.06}$ & ($-10\fdg3, 11\fdg2 $) & ($-65\fdg58,  65\fdg58 $)  \\
 &  & $ 2458683.84080_{-0.02004}^{+0.02007}$ & & & & & & \\
271892852 & 4.1245$\pm$0.0020 & $4.14677_{-0.00001}^{+0.00001}$ & 41.71$_{-1.17}^{+1.78}$ &  4.27$_{-0.07}^{+0.07}$ & 83.29$_{-0.32}^{+0.43}$ & 5.00$_{-0.08}^{+0.09}$ & ($ 23\fdg7, 57\fdg4 $) & ($-77\fdg44,  77\fdg44 $)  \\
 &  & $2458411.045386_{-0.00166}^{+0.00166}$ & & & & & & \\
326257590 & 6.0426$\pm$0.0066\tablefootmark{*} & $6.05179_{-0.00001}^{+0.00001}$ & 17.31$_{-0.10}^{+0.11}$ & 5.39$_{-0.01}^{+0.02}$ & 89.48$_{-0.52}^{+0.32}$ & 3.92$_{-0.01}^{+0.00}$ & ($-9\fdg8, 10\fdg9$) & ($-72\fdg65, 72\fdg65$)  \\
 &  & $2458955.76484_{-0.00106}^{+0.00109}$  & & & & & & \\
\hline
\end{tabular}
\tablefoot{
\tablefoottext{*}{DASCH data resulted in 6.050 days period from the 90-yr long dataset, see Fig.\,\ref{3262_rotper}.}
}
\end{table*}

We present models for three binaries, TIC\,235934420, TIC\,271892852 and TIC\,326257590, which have data from the TESS CVZs for about four years, in some instances continuously for about a year. Such datasets allow us to compare results originating from eclipse mapping together with time-series spot modeling of the whole light curves. Basic information for the three binaries is summarized in Table\,\ref{3stars}. The columns give the TIC number, rotational period from the available TESS data using the program package \texttt{MuFrAn} \citep{2004ESASP.559..396C},  orbital period and T$\mathrm{_0}$ from the eclipse fitting results, flux ratio between the components ($k_f$), ratio of the radii of the components ($R_1/R_2$), inclination, semi-major axis of the binary orbit in the units of the primary star radius ($a/R_1$), and finally, the scanned longitudes and latitudes on the primary by the secondary star. In this paper we call those minima that show evidence of spots as primary minima.

The three systems,  apart from their long-term datasets, are interesting from different perspectives. TIC\,235934420 have useful two-color ZTF measurements contemporaneously observed with some of the TESS data allowing us to derive spot temperatures independently from eclipse mapping for the same time interval. TIC\,271892852 has its scanned region by the secondary star at higher latitudes, therefore we may trace the effect of the differential rotation in the eclipse maps. Finally, more than a year of continuous TESS data of TIC\,326257590 made it possible to follow the slight changes in spot positions without interruption. 

\subsection{Multicolor results for TIC\,235934420}\label{2359_results}

\begin{figure}[h]
    \includegraphics[width=\columnwidth]{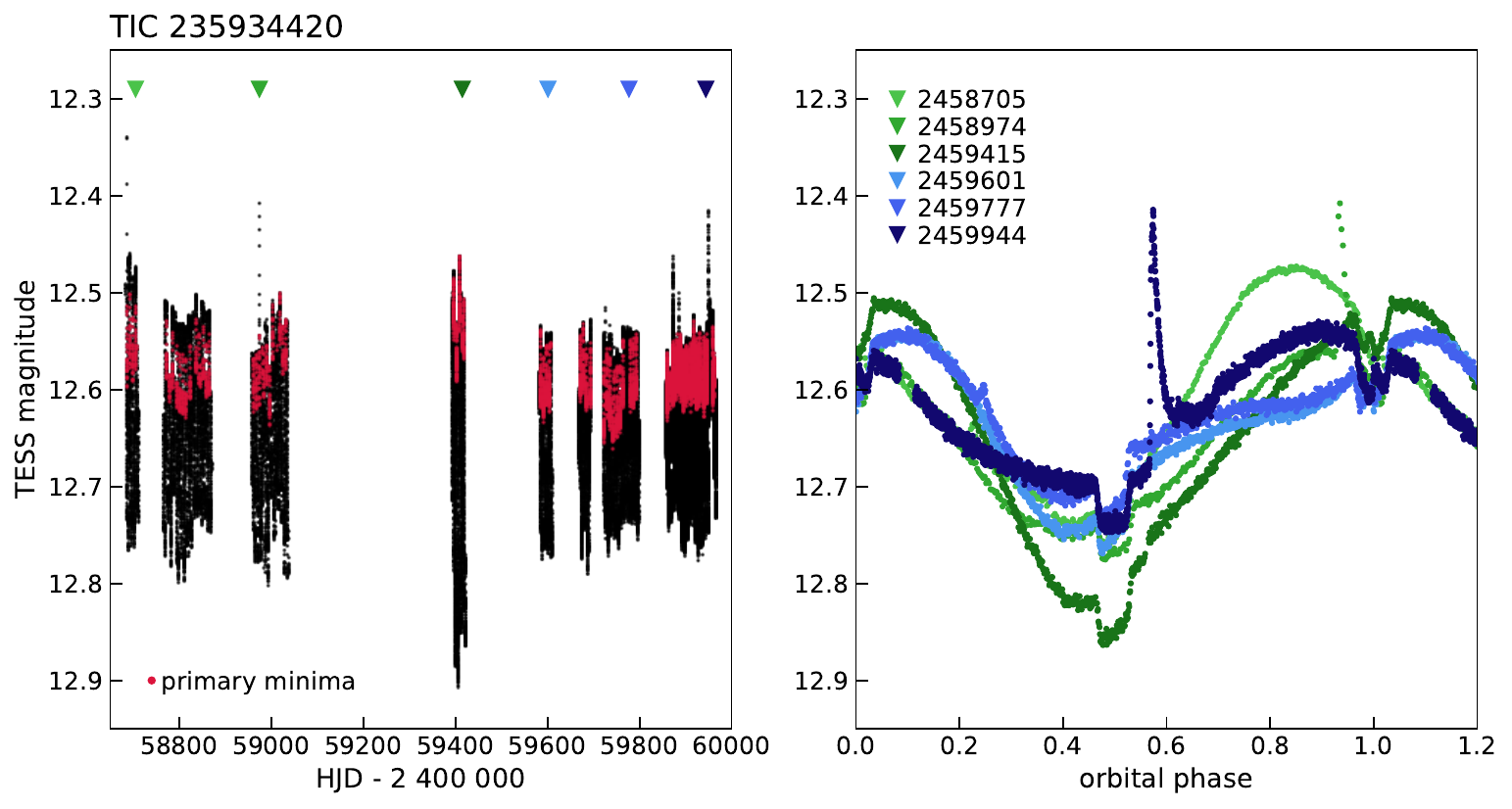}
      \caption{Left panel: TESS observations of TIC\,235934420 made between July 2019 and January 2023 with the primary eclipses marked with red. Six individual light curves are marked with different colors and are phased with the orbital period in the right panel, showing the variability of the light curves due to spots at different epochs. The large sharp excursions in flux are due to stellar flares.}
      \label{2359_lc_examples}
\end{figure}

\begin{figure}[h]
   \includegraphics[width=\columnwidth]{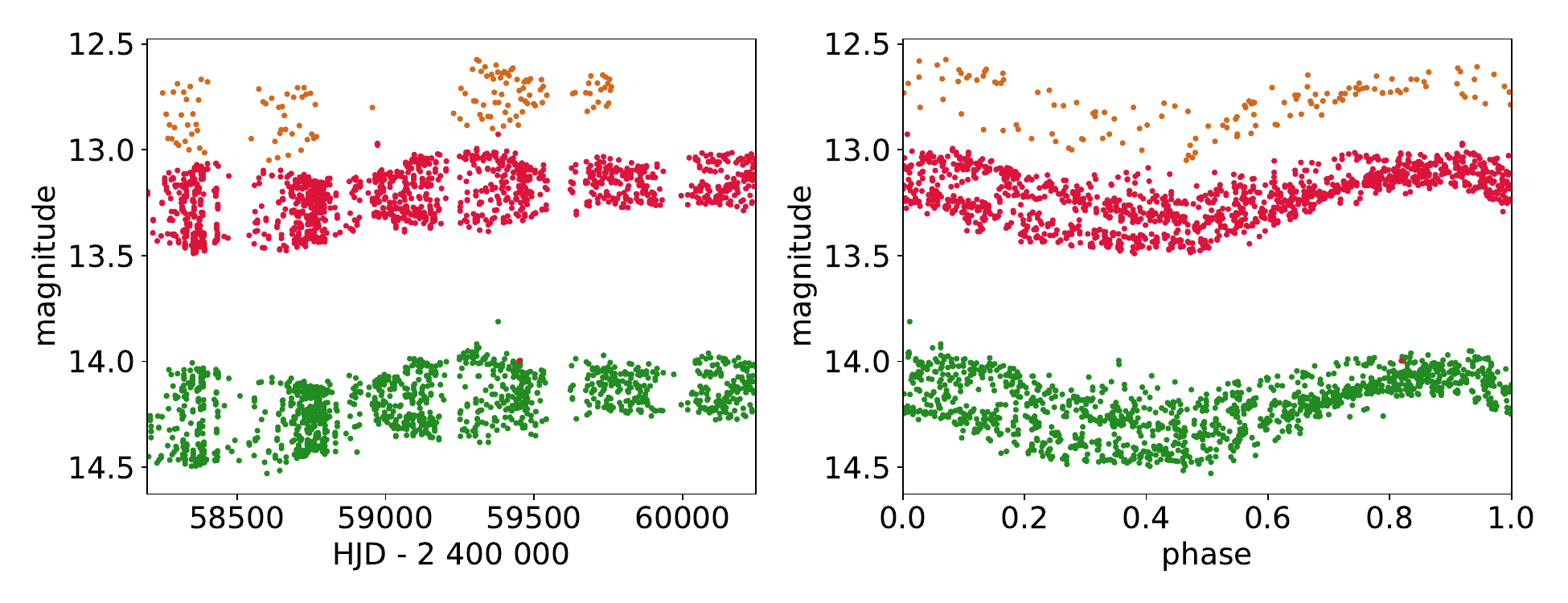}
      \caption{ZTF observations of TIC\,235934420. Green, red and brown points show the $g$, $r$ and $i$ band data, respectively. The shape and amplitude of the light curves are slowly and continuously changing.}
      \label{2359_ZTF}
\end{figure}

We use TESS observations made during 3.5 years between July 2019 and January 2023 shown in Fig.\,\ref{2359_lc_examples} with illustrative light curves at different epochs. Due to changes in spot activity we see variable light curves with typical modulation amplitudes of 0.2$-$0.3 magnitude. For most eclipses only one spot was recovered. 

ZTF provides approximately 2000 day long datasets to date, observed with well calibrated ZTF $g$, $r$, and $i$ filters of TIC\,235934420 which we present in Fig.\,\ref{2359_ZTF}. Part of the ZTF observations between HJD\,2458681-2458760 had dense enough phase coverage allowing us time-series spot modeling parallel in $g$ and $r$ filters yielding spot temperatures from the data. The details are given in Appendix\,\ref{ZTF_two-color}. From the results an average 3550$\pm 80$\,K spot temperature follows, which is used as constant spot temperature for all the TESS data in the time-series spot modeling. 

\begin{figure}[thb]
   \includegraphics[width=\columnwidth]{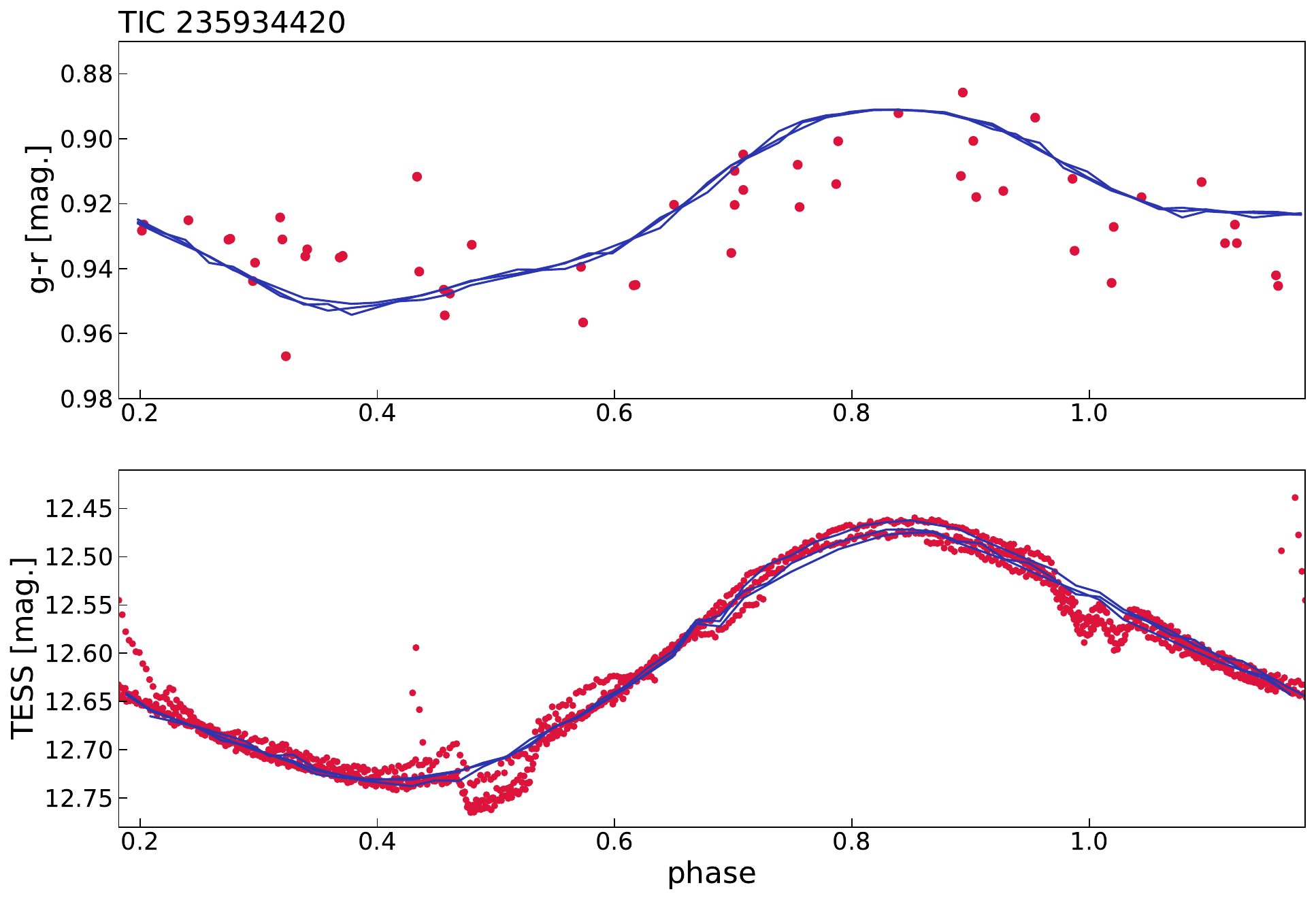} \includegraphics[width=\columnwidth]{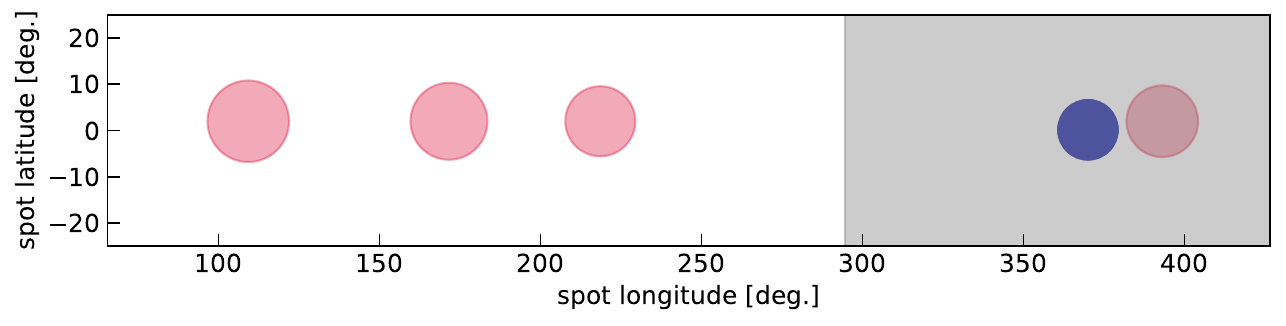}
      \caption{Contemporaneous ZTF and TESS data modeling of TIC\,235934420. Top: the fitted ZTF $g-r$ color index reflecting the temperature at the given rotation phase. The depressions in the color index curve around the derived spots indicate a redder color and thus a lower temperature.  Middle: phase folded fit to the contemporaneous TESS data. Bottom: locations of the spots on the stellar surface from the second observed minimum at HJD\,2458691.506  (see Fig.\,\ref{2359_tser}) from time series modeling (red) and eclipse mapping (blue). Grey shading shows the longitudes scanned by the companion star.}
      \label{2359_14_TESS}
\end{figure}

\begin{figure}[thb]
   \includegraphics[width=\columnwidth]{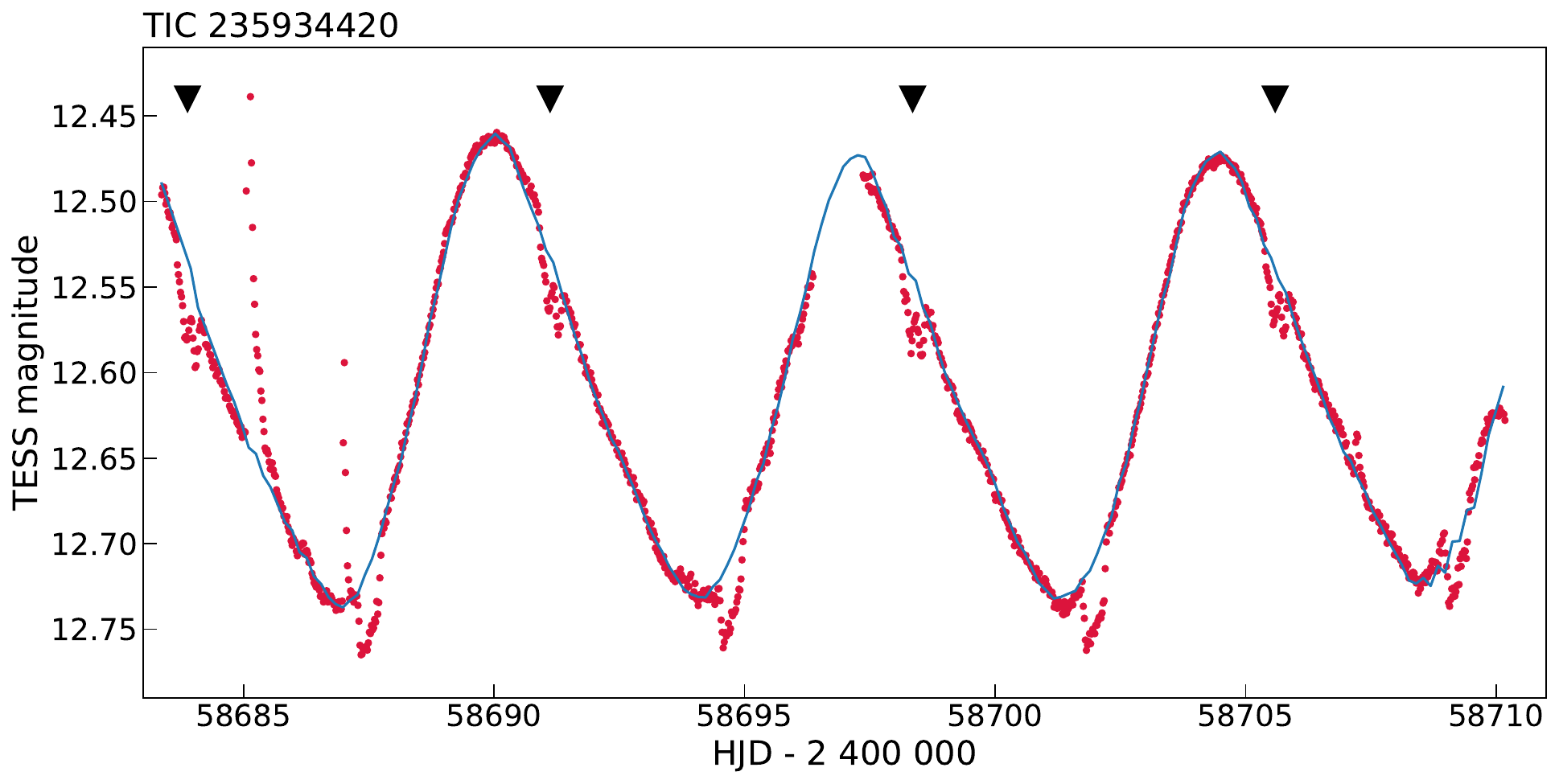}
      \caption{Time-series fits with four spots to the contemporaneous TESS data with ZTF to TIC\,235934420. The fit is drawn with gray line. The mapped eclipses are marked with arrows.
      } 
      \label{2359_tser}
\end{figure}

The comparison between the fits to the contemporaneous TESS (with 1800-s cadence only) and ZTF $g$ and $r$ data is seen in Fig.\,\ref{2359_14_TESS} folded with the orbital period. Only the smooth TESS light curve caused by spots is modeled, flares and eclipses are removed from the data. The upper panel shows the $g-r$ color index and its fit with three spots which is suitable given the precision and density of the data. The color index curves and their fits are simple subtractions from the individual $g$ and $r$ data and fits (Fig.\,\ref{2359_ZTF_tser}). The result also reflects the good calibration of the ZTF data. Below, in the middle panel, the full folded TESS data (eclipses and flares now included) with their fits are plotted, and in the bottom the spot configuration of the second eclipse from Fig.\,\ref{2359_tser} is seen. The redder color, thus lower temperature on the stellar surface around the spot longitudes is evident from the three panels of the figure.

\subsection{Results for TIC\,271892852}\label{2718_results}

\begin{figure}[h!!!]
    \includegraphics[width=\columnwidth]{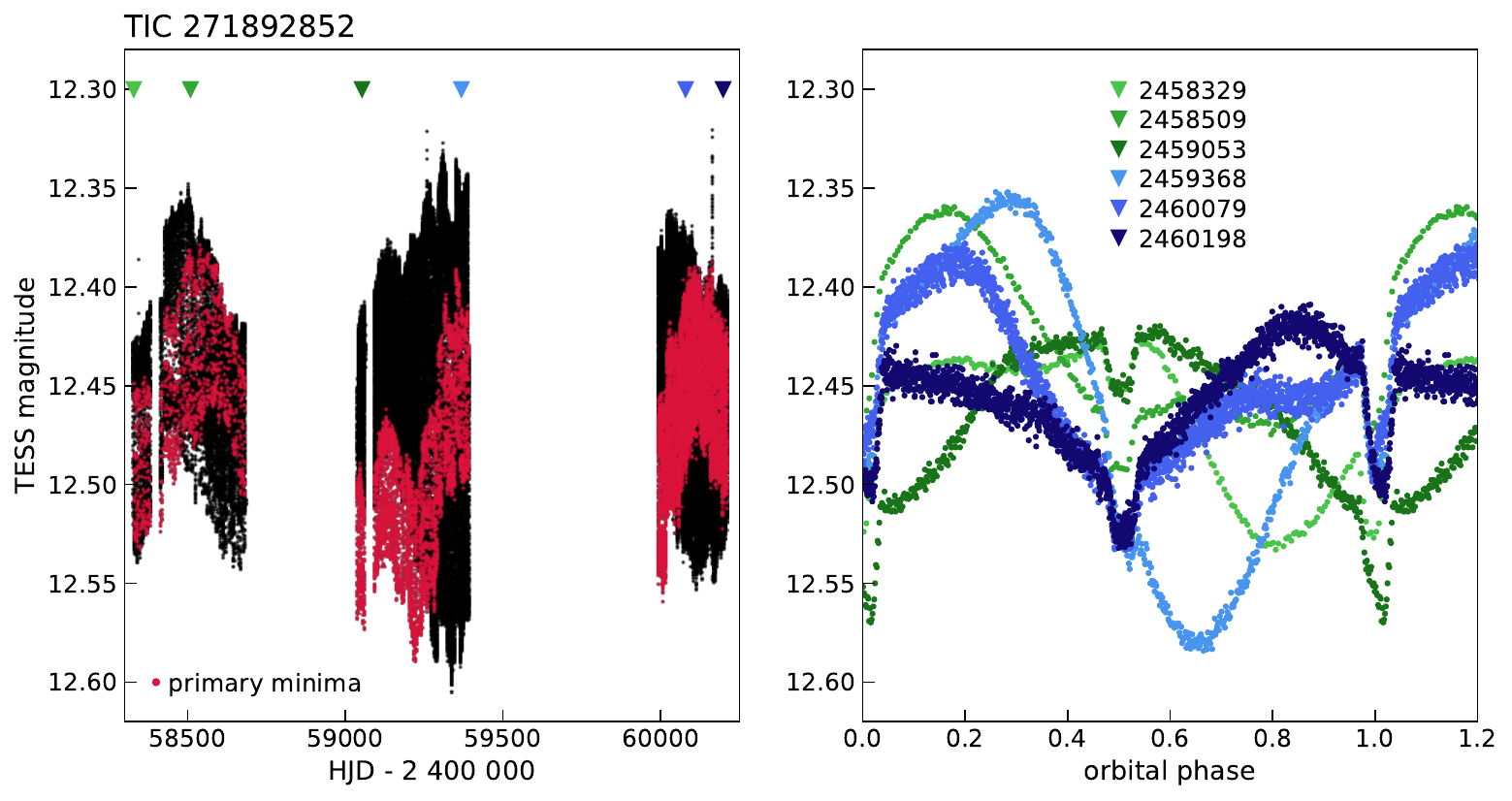}
      \caption{Left panel: TESS observations of TIC\,271892852 made between July 2018 and September 2023, with the primary eclipses marked with red. Six individual light curves are marked with different colors and are phased with the orbital period in the right panel, showing the variability of the light curves due to spots at different epochs.}
      \label{2718_lc_examples}
\end{figure}

\begin{figure}[h!!!]
    \includegraphics[width=\columnwidth]{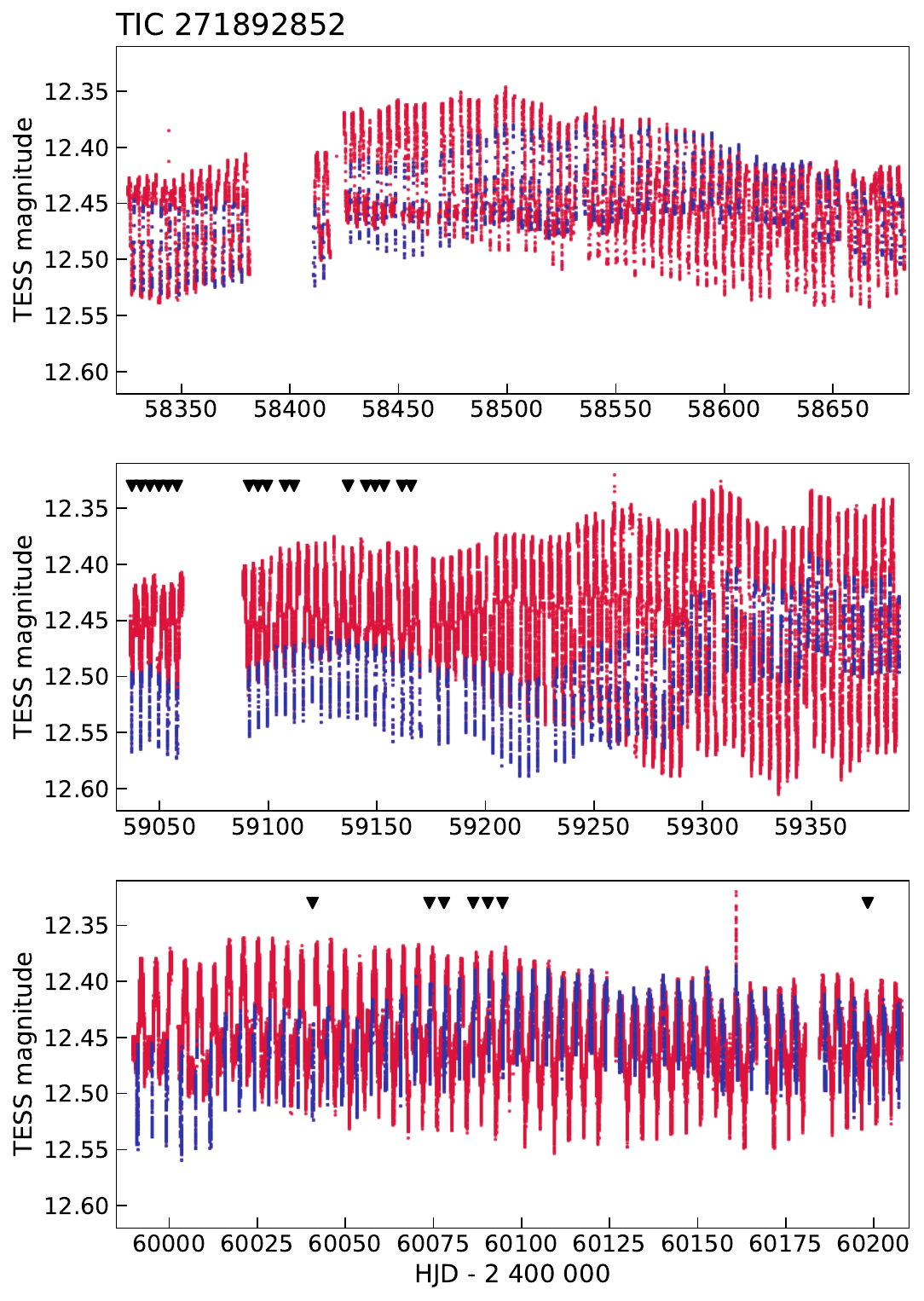}
      \caption{Three continuous sets of TESS observations. Blue points mark the primary eclipses. The systematic drift of the minima on the light curve indicates a difference between the orbital and the average rotational periods. See Sect.\,\ref{rotation} for more.
      }
      \label{2718_3parts}
\end{figure}

The TESS dataset of TIC\,271892852 consists of three, almost fully continuous datasets observed between July 2018 and September 2023. The light curve, plotted in Fig.\,\ref{2718_lc_examples} shows that its shape is strongly variable exhibiting continuous drifts of the primary minima. This is detailed in Fig.\,\ref{2718_3parts} depicting the variations in the 1800-s (top), 600-s (middle) and 200-s (bottom) cadence data. 

\begin{figure*}[thb]
\valign{#\cr
  \hbox{%
    \begin{subfigure}[b]{.5\textwidth}
    \centering
    \includegraphics[width=8cm]{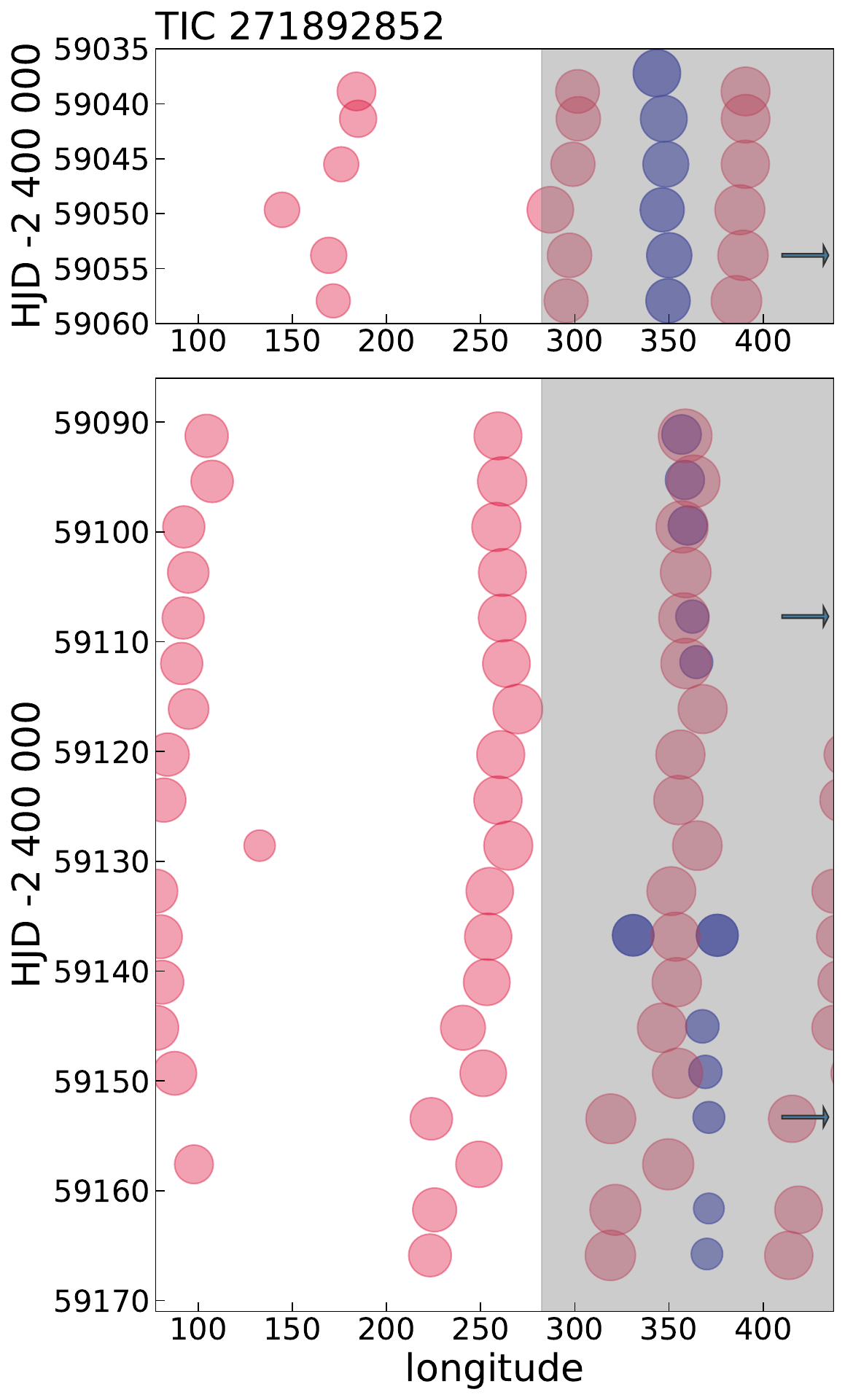}
  \end{subfigure}%
 }\cr
  \noalign{\hfill}
  \vspace*{0.3cm}
 \hspace*{-0.5cm}
   \hbox{%
    \begin{subfigure}{.5\textwidth}
    \centering
    \includegraphics[width=8.4cm]{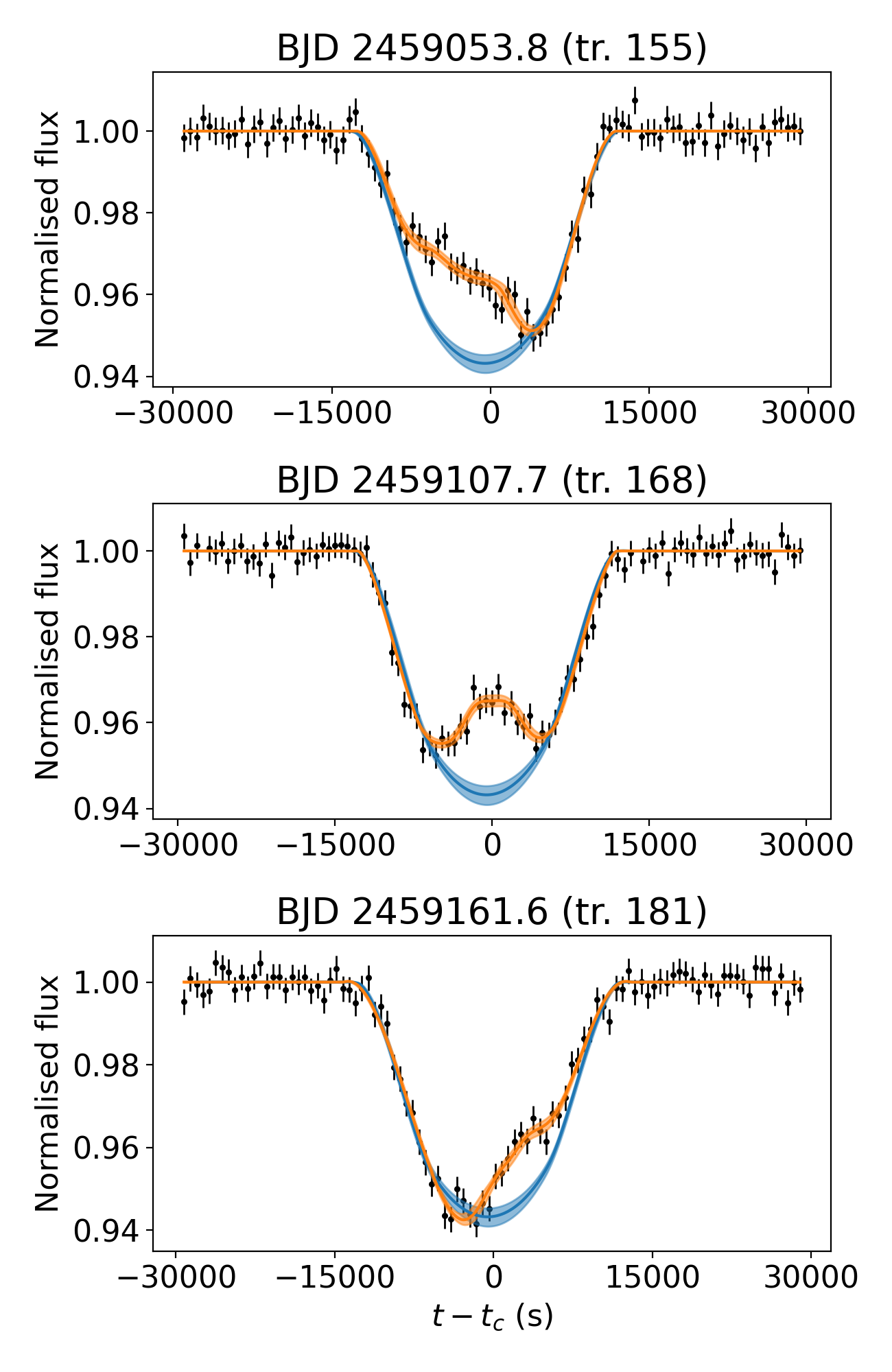}
      \end{subfigure}%
  }\cr
  }\vfill 
 \caption{Left: changing longitudes of spots of TIC\,271892852 in time. Red and blue circles result from time-series photometric modeling and eclipse mapping, respectively. The shade of the blue circles reflects the contrasts of the spots from eclipse mapping. Note that the x-axis of the figure covers one stellar circumference in longitude. Gray shading shows the longitudes scanned by the companion star. Three arrows mark the rotations with fitted spots whose eclipses are shown in the right panel. Right, from top to bottom: example eclipse fits revealing spots scanned already during ingress, just in the middle of the eclipse, and during egress.}   
 \label{2718_spots}
\end{figure*}

A series of eclipse maps was calculated in the beginning of the 600-s cadence data (Fig.\,\ref{2718_3parts}) during about 130 days when the primary minima occurred near the light curve minima. From the eclipse maps in most cases only one spot was recovered as seen in the left panel of Fig.\,\ref{2718_spots}; the data have good time resolution but the fluxes have considerable scatter due to the faintness of the star. The extraction of the minima were successful mostly where the light curve did not change too much and/or too differently before and after the minima. 

Interestingly, the scanned latitudes of this binary are high, and the spot latitudes from the eclipse fits vary from $37\fdg5$ to 50$^{\circ}$. The average is $\approx$40$\pm$17$^{\circ}$ (cf. Table\,\ref{3stars}) which we used as a fixed latitude for modeling the whole light curve. Derived spot latitudes from photometry also tend to prefer high latitudes when are set as free parameters, as our experiment shows in Appendix\,\ref{spotlatitudes}.

A continuous drift of spot longitudes from time-series modeling of the full light curves is evident on the left side of Fig.\,\ref{2718_spots}, probably due to differential rotation, at least partly. The right side of the figure shows three eclipse fits as examples. These fits show a changing eclipse feature from partial on the ingress side through a total in the middle to again a partial on the egress side. The baseline of the transit may change, as transits from different TESS sectors were sampled separately. We show an average baseline and its scatter (in blue) calculated from all the fits from the sectors of the three displayed examples.

\subsection{Results for TIC\,326257590}\label{3262_results}

\begin{figure}[thb]
    \includegraphics[width=\columnwidth]{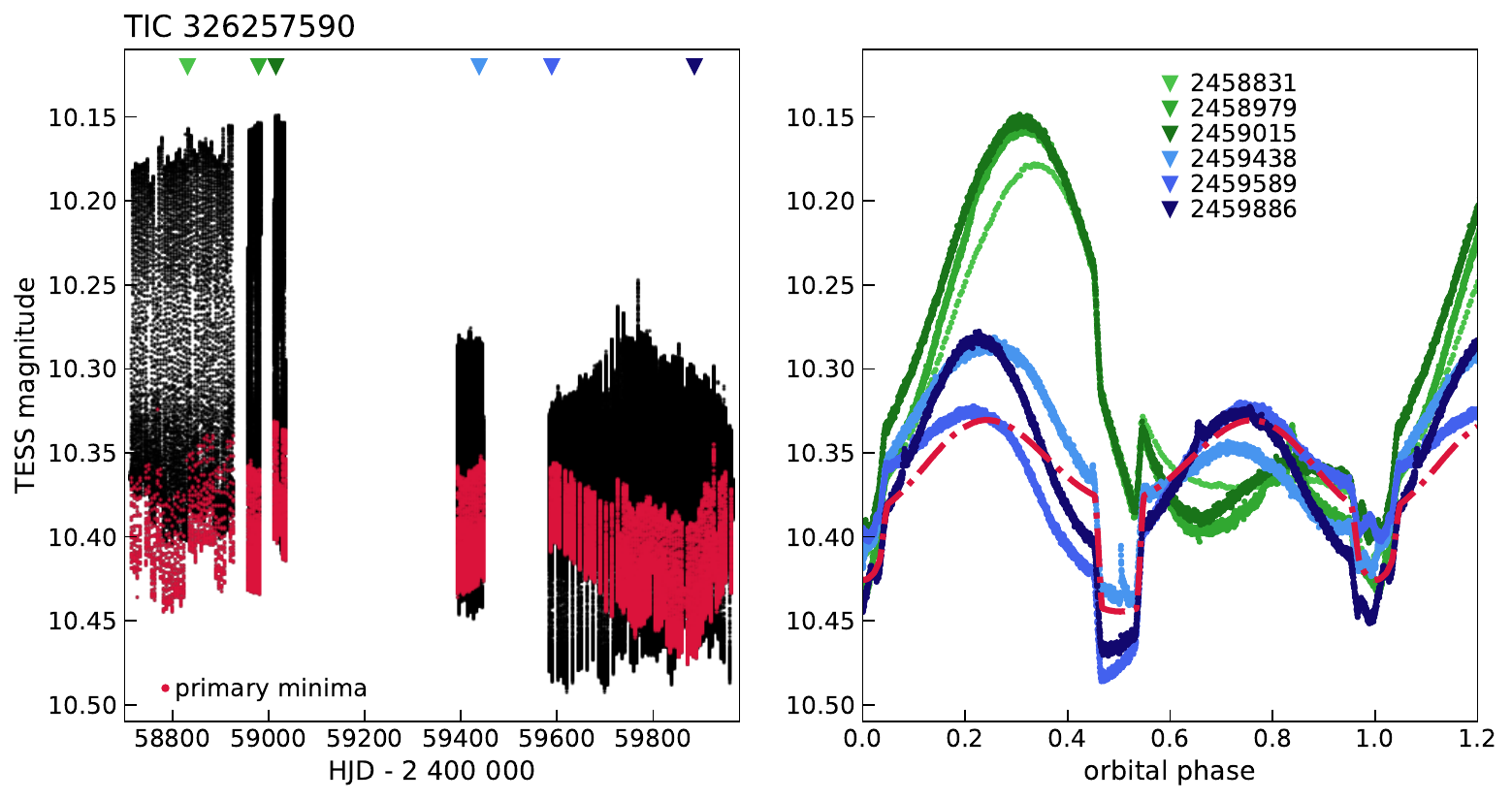}
      \caption{Left panel: TESS observations of TIC\,326257590 made between August 2019 and January 2023, with the primary eclipses marked with red. Six individual light curves are marked with different colors and are phased with the orbital period in the right panel, showing the variability of the light curves due to spots at different epochs. Red dash-dotted line shows the approximate ellipsoidal light variations shifted to the most dense part of the observations for comparison.}
      \label{3262_lc_examples}
\end{figure}

Long, continuous datasets have been observed by TESS for TIC\,326257590, which can be used for tracking spots on the stellar surface. The example light curves seen in Fig.\,\ref{3262_lc_examples} together with the detected 100 flares (see Fig.\,\ref{fig:example_lc}) bear witness to the strong variation in the activity. Atmospheric parameters, rotational and radial velocities, galactic positions and space velocities of this binary are given in \citet{2020A&A...637A..43K}, where the presence of a strong Li line is also noted. 

\begin{figure*}[thb]
    \includegraphics[width=7.75cm]{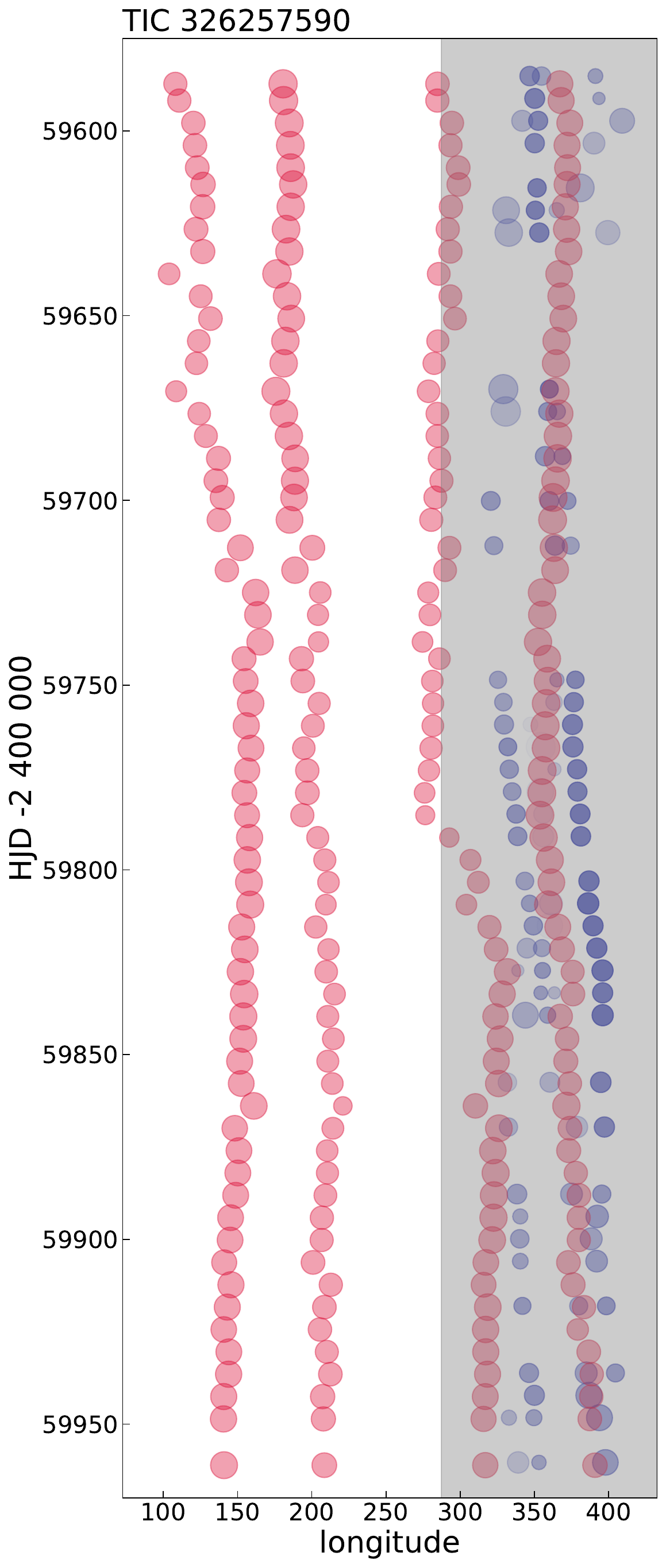} \hfill 
    \includegraphics[width=9.8cm]{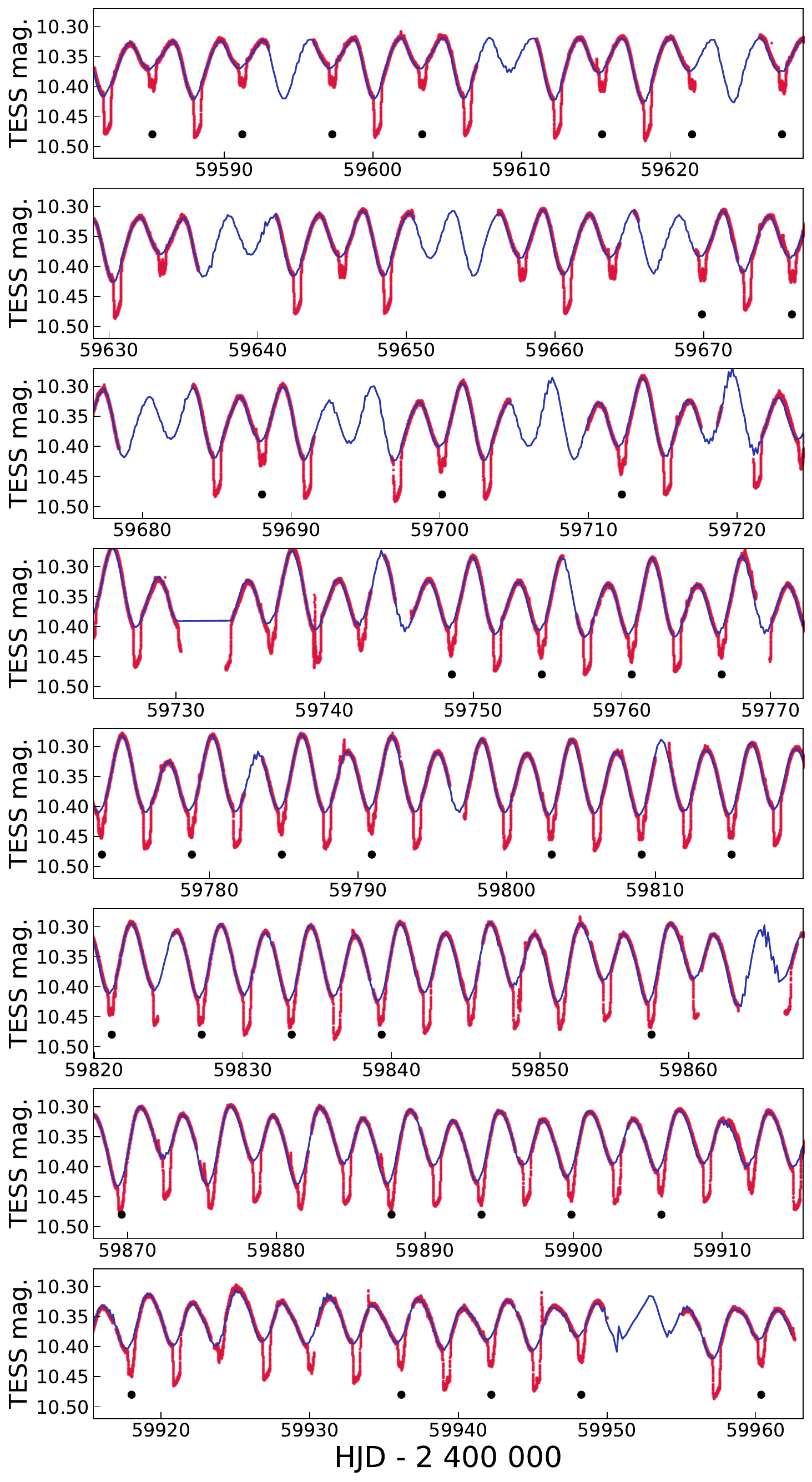}
      \caption{Left: Changing longitudes of spots of TIC\,326257590 in time. Red and blue circles result from time-series photometric modeling and eclipse mapping, respectively. The shade of the blue circles reflects the contrasts of the spots from eclipse mapping. Note that the x-axis of the figure covers one stellar circumference in longitude. Gray shading shows the longitudes scanned by the companion star.  Right: Light curve fitted by four spots, black dots mark those eclipses which are mapped.}
      \label{3262_endfits}
\end{figure*}

The observations of this object date back to more than 120 years plotted in Fig.\,\ref{3262_rotper}, which show only small long-term changes in the light curves without any sign of cycles on a decadal timescale. The DASCH, ASAS-SN and TESS datasets have very different sampling, but the rotational modulation caused by spots is already obvious from the 90 years of sometimes incomplete photographic dataset. The ASAS\,SN observations show a small secular variation, and half of the rotational period appears as well. Only the TESS data revealed that the object is an eclipsing binary. For a period search of the TESS data we used the clean light curve free from eclipses and flares, and the 120-s and 200-s data down-sampled to every 10th datapoint. In the amplitude spectra of the ASAS and TESS data multiple peaks mark the rotational period as possible sign of the differential rotation (see also Fig.\,\ref{diffrot}), but it is out of the scope of the present paper to further investigate this.

\begin{figure*}[thb]
   \centering
   \includegraphics[width=16cm]{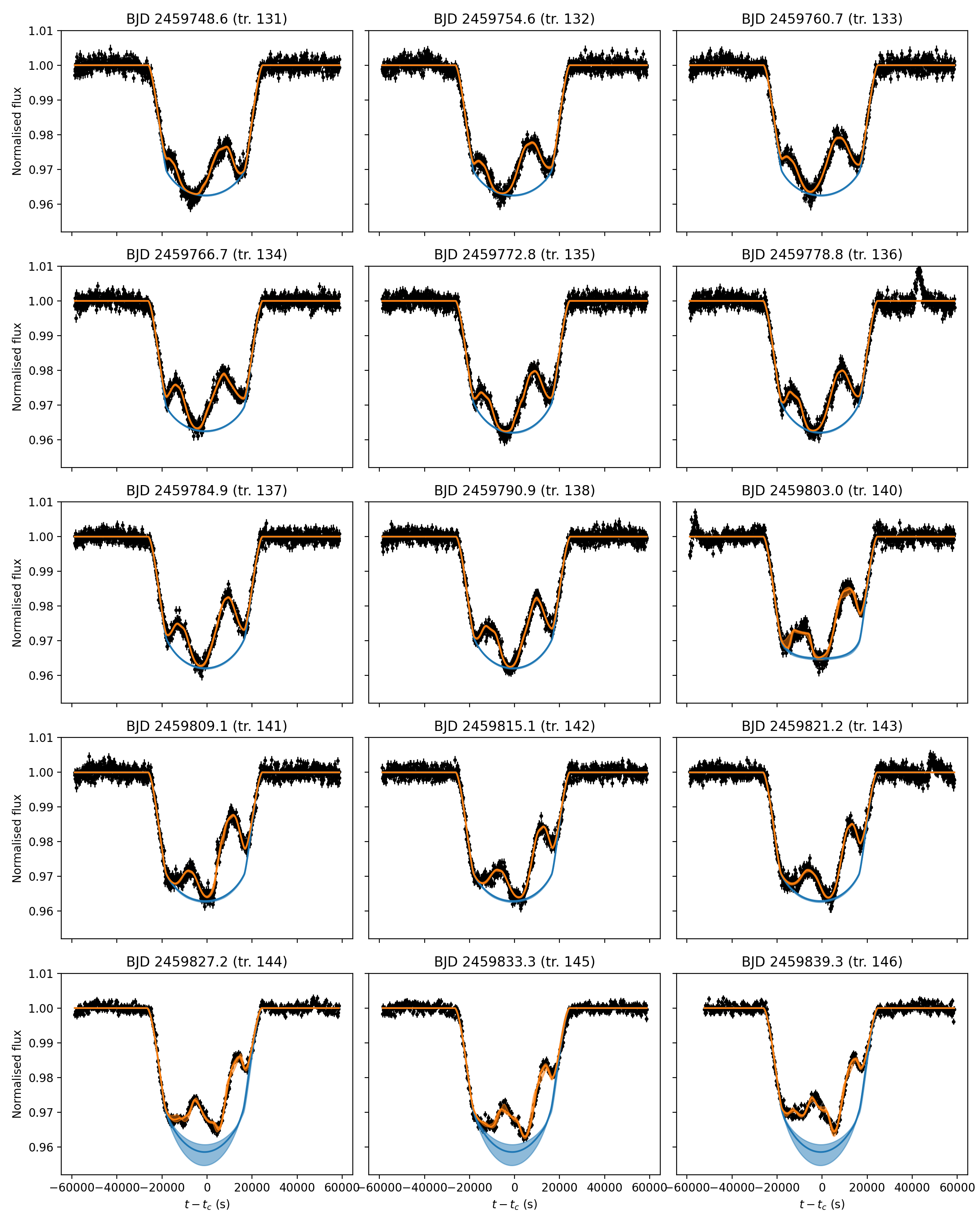}
      \caption{Examples of fits in a continuous set of eclipses for TIC~326257590 between HJD 2459748-245933. The transit base line may have bigger uncertainties if the transit contains several long spot eclipses, as seen in the final row.}
      \label{fitfigure}
\end{figure*}

More than a year long (381 days) continuous TESS observations  allowed us to monitor the changes of the spots' longitudes via time-series modeling of the full light curves (eclipses and flares removed) and through eclipse mapping on an approximately 145$^{\circ}$ ($\sim$0.4$P_{\rm orb}$) wide region around the substellar point. Altogether 37 eclipses were successfully mapped during 63 rotations/orbits. The eclipse maps uncover usually three (a few times only two) spots in every scan from the accurate and high time-resolution data which were 120-s before HJD 2459822 and 200-s afterwards. The results are plotted Fig.\,\ref{3262_endfits}, where the full light curves and the spot positions can be identified by their Julian dates. A nearly uninterrupted sequence of fitted eclipses are plotted in Fig.\,\ref{fitfigure} as examples.

From the four suspected spots from the time-series photometric spot modeling two appear around the substellar point, and the eclipse mapping results show two to three spots in the same region. On the opposite side of the primary two other spots are revealed by the time-series modeling. Small systematic drifts of the spots are found to be very similar in the two methods during most of the time, as can be followed in Fig.\,\ref{3262_endfits}. Experiments about spot latitudes using the data of very high accuracy suggest that there could be higher latitude spots on the other side of the giant star. Details of this experiment are given in Appendix\,\ref{spotlatitudes}, and in Fig.\,\ref{3262_fits}. 

\section{Discussion}\label{sect_disc}

\begin{figure*}[t]
    \includegraphics[width=9.0cm]{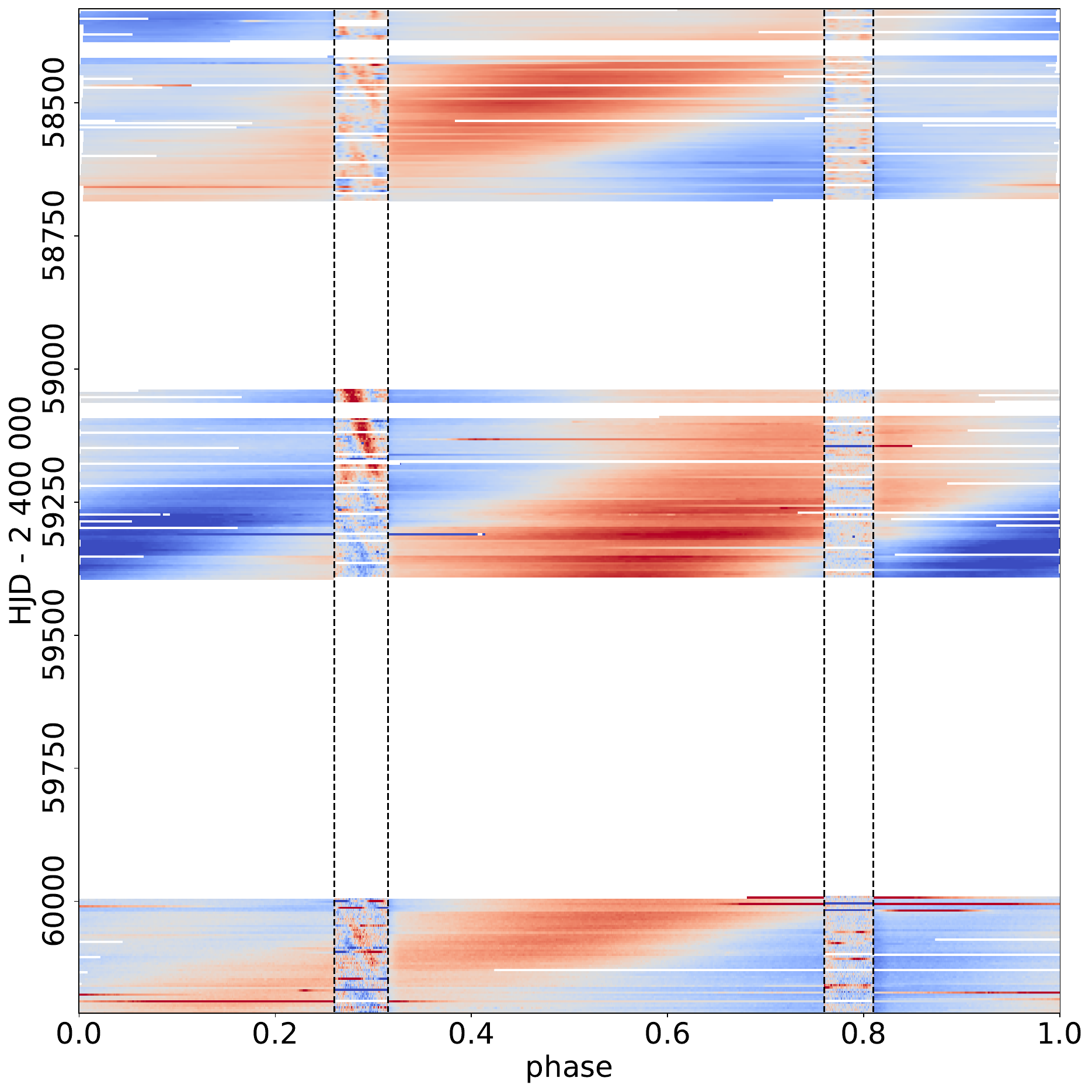} \hfill 
    \includegraphics[width=9.0cm]{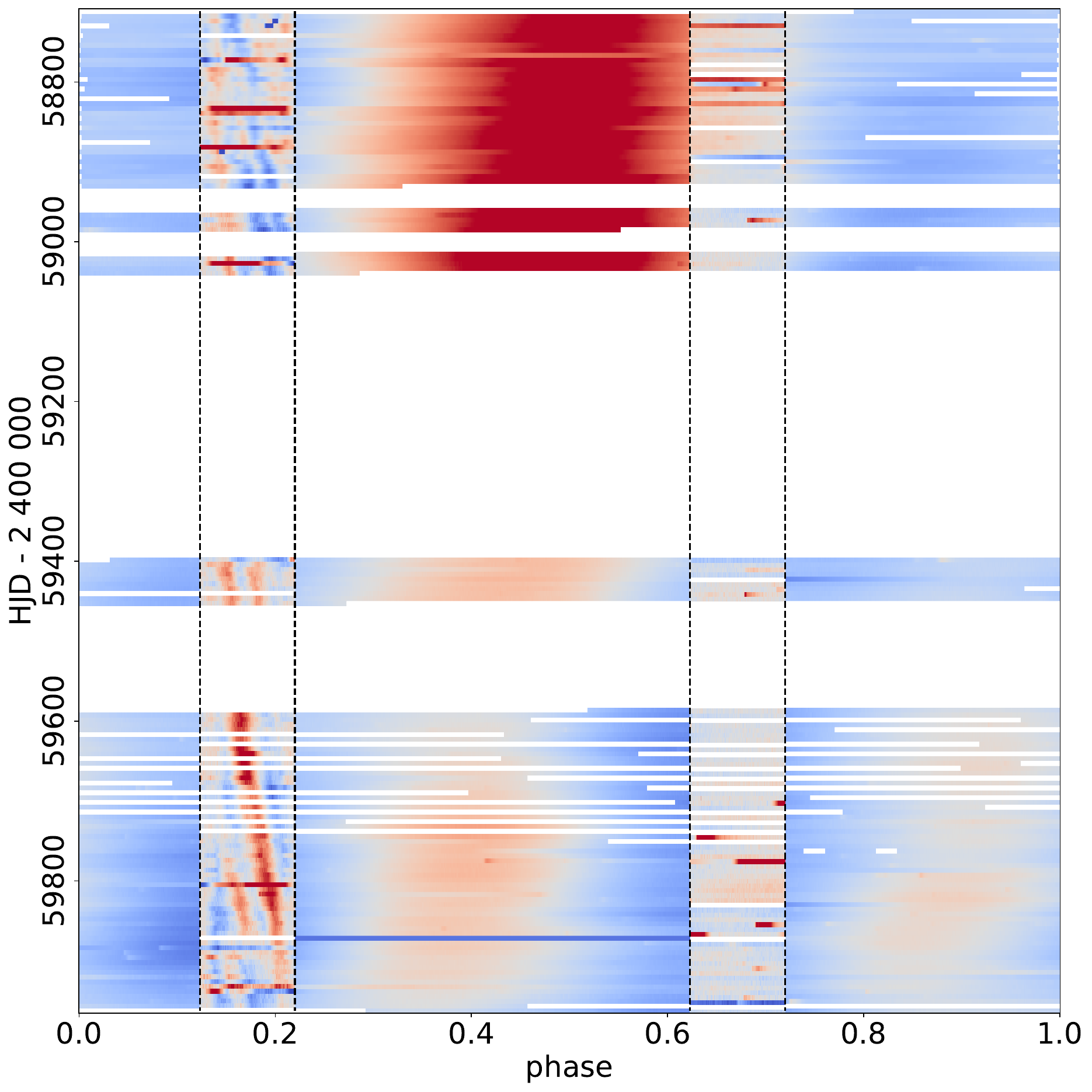}
      \caption{Full light curve of TIC\,271892852 (left) and TIC\,326257590 (right) folded with the orbital periods of the systems. In the out-of-eclipse phases, the original light curve is shown, illustrating the spot modulation, whereas during the eclipses (marked with dashed lines) the residual light curve is shown after removing the average eclipse signal, enhancing the spot occultation features. The color coding goes from blue to red with increasing brightness, scaled separately in- and out-of-eclipse to highlight the main features.
      }
      \label{full_maps}
\end{figure*}

In Fig.\,\ref{full_maps}, we show phase folded light curves for TIC~271892852 and TIC~326257590, as two dimensional maps. We fold the light curves with the orbital period, and plot them as a function of phase and time, color coded with flux, thus showing the spot modulation pattern. During the eclipses (marked with dashed lines), we plot the eclipsing light curve residuals instead, subtracting the averaged eclipses to reveal the spot occultation pattern. White regions indicate missing data or incorrectly extracted eclipses. It can easily be seen that the orbital and rotational periods slightly differ, causing spots to shift in phase in the reference frame of the orbit. Due to the scanning of the secondary star, the slopes of the spot shifts are mirrored during the eclipses and the out-of-eclipse modulation. More details are given in Sect.\,\ref{rotation}.

All three binaries studied in this paper have spots on their substellar regions. As an example, such a feature appears on the giant primary of the binary UZ\,Lib where the positions of equatorial spots point towards the secondary star and opposite to it, as have been found by \cite{2002A&A...389..202O} from Doppler imaging.

\subsection{Differential rotation/spot positions}\label{rotation}

\begin{figure}[thb]
   \includegraphics[width=\columnwidth]{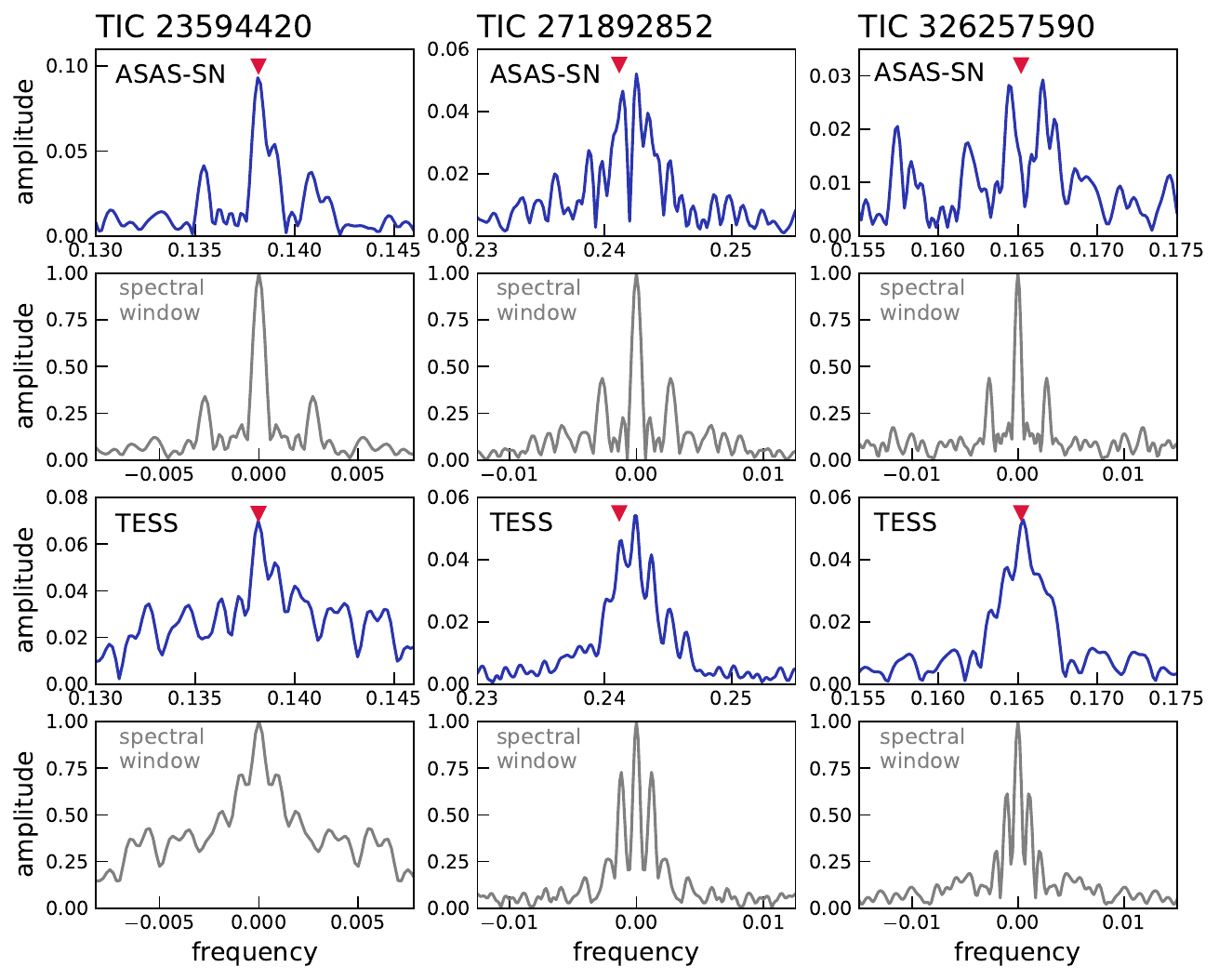}
      \caption{Amplitude spectra with spectral windows around the rotational frequencies of the three giants from ASAS\,SN and TESS data. The spectral windows are shown in gray. Red arrows mark the frequency of the binary orbit.}
      \label{diffrot}
\end{figure}

The three studied systems are circularized and synchronized (at least they are close to that) which is determined from their light curves showing very similar orbital and rotational periods, and from the fact that the two eclipses happen half of the orbital period from each other. On Fig.\,\ref{diffrot} the amplitude spectra of the three giant primaries are plotted and these typically exhibit two or more significant peaks. Red triangles mark the orbital frequencies. The reason for these multiple peaks could be differential rotation, but drifts in spot locations \citep[see e.g.][]{2000A&A...354..537S}, and/or new spot appearances at similar longitudes could also play a role when we observe slightly changing rotational periods. 

For TIC\,23594420 and TIC\,32627590 the position of the orbital frequency is close to the middle of the rotational peaks which means that the observed rotational periods at different times are slower and faster compared to the orbital ones by about the same amount.  The orbital period of TIC\,271892852 and the rotational period of the primary star are also close to each other, but there is a small, significant difference between them, the rotational one being slightly slower most of the time.  The global rate of the drifting spot longitudes of the three stars are  0.03, 0.38 and 0.09$^\circ$\,day$^{-1}$ for TIC~235934420, 271892852, and 326257590, respectively.

The existence of spots on TIC\,271892852 at higher latitudes around +40$^{\circ}$ is demonstrated by eclipse maps where the secondary star scans the primary, and lower latitude spots could also be present (cf. Fig.\,\ref{2718_fits}). Figure\,\ref{full_maps}, left, shows the drift of the light curves with respect to the orbital period. This difference could be due to a small, antisolar type differential rotation, but cannot be verified. The different longitudinal  drifts of the spots with respect to each other on TIC\,32627590, shown on Fig.\,\ref{3262_endfits} and Fig.\,\ref{3262_fits}, also point toward the existence of differential rotation.

On the differentially rotating stellar surfaces, at some given latitudes, the orbital period of the binary is equal with the local rotational period. To find that special latitude theoretically is a complicated problem and subject of many different stellar and orbital parameters as studied by \citet{1981ApJ...246..292S, 1982ApJ...253..298S}.  We find spots near the equator to high latitudes on our three binaries. Unfortunately, not knowing the corotation latitudes we cannot find the sign of the differential rotation for sure from our data.

Generally, not much observational evidence exists about the size and sign of the differential rotation in the different types of stars, especially in binaries with active components. However, surface differential rotation of late-type active (sub)giant stars is usually found to be smaller than that of the Sun \citep[e.g.,][etc.]{2004MNRAS.348.1175P,2004MNRAS.351..826P,2007AN....328.1075W,2007AN....328.1047M}. Given that rapidly rotating active giants are often members of RS\,CVn type close binary systems, their weak differential rotation is also consistent with the fact that tidal coupling in such systems is expected to diminish differential rotation \citep[cf.][]{2016A&A...593A.123O,2023A&A...674A.143K,2024ApJ...976..217X}.
For a recent review of known differential rotations based on Doppler imaging results of about a dozen giants/subgiants in binaries see \citet{2017AN....338..903K}. Table~1 of this paper also shows that an antisolar type of differential rotation is exclusively observed on subgiant/giant stars.

Scanning part of the surface of the giant star by its companion, determined by the binary orbit, gives good estimates of spot latitudes. Modeling the whole light curve using the latitudes that are safely known from eclipse mapping, we find spots near the same positions in the scanned region. The two approaches (eclipse mapping and full light curve solution) strengthen each other for this small part of the surface of the giant. The high amplitudes of the full light curves indicate the presence of the spots outside the eclipse chord as well, whose latitudes are uncertain. 

As to our present knowledge, starspots appear where the magnetic flux tubes reach the surfaces of the stars. This mainly depends on the rotational velocity, the depth of the convection zone and the strength of the magnetic field, shown by \citet{2007MmSAI..78..271H}, and more recently discussed by \citet{2024ApJ...976..215I}. In the case of binaries the emergence latitude could be modified by the gravitational force of the companion star. For dwarf stars in binaries \citet{2003A&A...405..303H} theoretically investigated the flux tube eruption, thus surface distribution of spots, taking into account different strengths of magnetic fields and orbital periods. For the much more diverse group of giant stars in binaries, as to our knowledge, no such simulation exists. The present results together with the study of the remaining binaries from the whole sample (Table\,\ref{basic_data} and Haris et al. in prep.) may give some observational constraints for such a study in the future. 

\subsection{Spot temperatures/contrasts/sizes}

As explained in Sect.\,\ref{ts_spotmodel}, for the time-series modeling of the full light curves the resulting spots are those which describe the light variations without taking into account the long-term changes of the brightness (i.e., for the unspotted brightness we use the local maxima for each dataset in the time-series modeling). The derived spot sizes are similar to each other from the eclipse mapping and full light curve modeling on the part of the stellar surface scanned by the secondary star, and are also similar outside that region, for all three giants. This similarity suggests that the spot contrasts (temperatures), which are kept constant in the course of the time-series modeling of the full light curves, are indeed well chosen. In the scanned regions of the giants the spots' radii are typically between a few to about 25$^{\circ}$ both from eclipse mapping and spot modeling.

Luckily, we had contemporaneous observations by TESS and ZTF for four orbits of TIC\,235934420 as shown in Sect.\,\ref{2359_results}, which made it possible to compare the spot temperature results from the ZTF two-color $g-r$ data (for the details see Appendix\,\ref{ZTF_two-color}) and that from the eclipse mapping. From the two-color ZTF data for spot temperature we obtained $3550\pm80$\,K while from the eclipse mapping the spot temperatures calculated from the median of the resulting contrast values are about 3600\,K, 3500\,K and 3700\,K from the 1800-s, 600-s and 200-s data, respectively. For most eclipses one spot was recovered. Considering the uncertainty of both processes of the order of 100\,K, these values can be taken as reassuringly similar values. 

Based on this finding for the full light curve modeling we used fixed spot temperatures resulting from the eclipse maps for all three giant stars. In the case of TIC\,235934420 the fixed spot temperature for the light curve modeling was 3550\,K. From the resulting mostly single spot of the eclipse mapping, we got 3700\,K as a spot temperature of TIC\,271892852. TIC\,326257590 is a bright star with accurate data; the eclipse maps revealed at least two, but mostly three spots with different spot temperatures during the eclipses, therefore a general mean spot temperature has higher uncertainty. The median spot temperature was 3500\,K which served as constant spot temperature for the light curve modeling.

\begin{table}[thb]
\small
\centering
\caption{Resulting spot temperature, radius and coverage values}
\label{3stars_temp}
\setlength{\tabcolsep}{4pt} 
\begin{tabular}{cccccc}
\hline\hline                
\noalign{\smallskip}
TIC       &   $T_\mathrm{star}$ &   $T_\mathrm{sp}$\tablefootmark{1} & $T_\mathrm{sp, cal}$\tablefootmark{2} &  average spot radii &    coverage  \tablefootmark{3}     \\         
    No.      &    [K]   &     [K] & [K] &  [$^{\circ}$]   &   \\
\hline\noalign{\smallskip}
235934420 &   4482   &   3550  &  3537 & 17.7, 24.8, 21.8, 16.7 &  12.7 \\
271892852 &   4873   &   3700  &  3719 & 20.0, 19.4, 21.4 &  9.3  \\
326257590 &   4526   &   3500  &  3558 & 19.9, 17.3, 19.6, 18.8 &  10.8  \\
\hline
\end{tabular}
\tablefoot{
\tablefoottext{1}{Average values based on eclipse mapping results, which are used in the full light curve modelings. For TIC\,235934420 the value is verified by ZTF data, see Sect.\,\ref{ZTF_two-color}}. \tablefoottext{2}{{Calculated based on \citet[][see their Eq.\,4]{2021ApJ...907...89H}}. \tablefoottext{3}{Spot coverage in percent of the total stellar surface.}}}
\end{table}

Checking our values we applied Eq.\,4 from \cite{2021ApJ...907...89H} based on earlier spot temperature results. We used the stellar temperatures from Table\,\ref{basic_data}, and the results are given in Table\,\ref{3stars_temp} showing excellent agreement. The three similar giant stars of our sample have similar spot coverage in percent of the total stellar surfaces, and the spots are about 800-1000\,K cooler than the respective stellar temperatures.

An experiment with 3500\,K and 3700\,K spot temperatures for TIC\,271892852 shows only 1$-$2$^{\circ}$ difference in the resulting spot radii, while the deviation between the spot coordinates are negligible. In addition, one should keep in mind that what we call spots are rather activity complexes, whose average temperatures depend on the  umbra/penumbra/facula ratios, which can change with time, this way increasing the uncertainty of the spot radius results from using just one fix spot temperature.

\subsection{Evolutionary states}

Some of the binaries from our sample are located redward and below the subgiant branch of the Hertzsprung--Russell diagram (HRD) shown in Fig.\,\ref{HRD}. Such positions of active binary stars have already been found in the old open cluster M67 \citep{2003AJ....125..246M} and in other open and globular clusters \citep{2017ApJ...840...66G}. \citet{2022ApJ...927..222L} extended the search for active RS\,CVn binaries to the galactic field, and provided a catalog of nearly 500 such systems. It may well be that some of the binaries of this sample also belong to these special systems, which could be in a particular phase of their evolution, but the nature of this status is not understood yet. Recently, \cite{2025arXiv250405561D} identified 38 RS\,CVn stars as sub-subgiants, positioned on the HRD where a few of our systems are also found (see Fig.\,\ref{HRD}), concluding that they are "generally overactive short-period synchronized RGB binaries". Additionally, from the compilation of \citet{2022A&A...659A...3S}, some of the 17 highly active, evolved stars (mostly binaries) having strong chromospheric and coronal activities and high rotational velocities, are located also to the red and below the subgiant branch on the HRD, as are the above mentioned systems, see Fig.\,10 of the cited paper.

The positions of the stars on the HRD in most cases reflect only a snapshot of their life cycle. Systematic monitoring of the whole sky, or at least some part of it, dates back not much longer than a century. More accurate photometry which is necessary to precisely position the stars in the HRD are at best a few decades long. However, on a long time-scale, the brightness and color index changes of active binaries could be beyond 1.0 and 0.3 magnitudes, respectively \citep[cf. XX\,Tri,][]{2014A&A...572A..94O}. The reason for such a large brightness and temperature change ($\approx$300\,K, derived from the color indices) over a half-century long time-scale could hardly be explained fully by magnetic activity. For a longer discussion of this issue and, in particular, the possible global impact of flux blocked by large cool starspots, see \citet{2024NatCo..15.9986S}. In any case, activity can generate an extra scatter into the subgiant branch roughly in the direction of the interstellar reddening.

\section{Summary}

In the present pilot project we intended to use the advantage of binarity in studying the surface activity of giant stars. Having the active giant in an eclipsing system allows us to gather significant information about the stellar surface structure that otherwise is impossible from only photometric data. 

\begin{itemize}
 \item We gathered a sample of 29 eclipsing binaries with active subgiant/giant components, most of them showing bumps in their primary eclipses which are identified as starspots scanned by the secondary stars.
 \item The eclipse mapping technique, which was originally developed for planets transiting their host stars, is modified to deal with two stars, to extract the starspot parameters on the active primaries of eclipsing binaries.
 \item As a pilot study we present the analysis of the synchronized close binaries TIC\,235934420, TIC\,271892852 and TIC\,326257590 from the two CVZs of TESS. Both the primary eclipses and the full light curves were modeled and the results compared. We found that spots are always present at the substellar points. 
 \item Orbital and rotational periods were derived for the three studied binaries, and indications of differential rotation were found.
 \item With the help of contemporaneous two-color ZTF and TESS observations of TIC\,235934420 we verified the spot temperature resulting from eclipse mapping observed by TESS as an average of 3600$\pm$100\,K with simultaneous full light curve fitting in $g$ and $r$ colors resulting in 3550$\pm$80\,K. 
 \item We show direct evidence of high latitude spots on TIC\,271892852 from eclipse mapping, since the middle of the transit chord of the secondary star is at 40$^{\circ}$ latitude. This technique is the only way to find starspot latitudes with high confidence from photometry alone. 
 \item On all three studied active giants we find slow, systematic changes of spot positions in time. These changes could partly be due to differential rotation, but may also be a sign of vanishing and newly emerging spots at similar longitudes, or spot proper motions.
\end{itemize}

In the continuation of this work (Haris et al. in prep.) we will present the analysis of about two dozen eclipsing binaries with active giant primaries from Table\,\ref{basic_data}, which are quite similar to each other, and are in similar evolutionary stage shown in Fig.\,\ref{HRD}.  The strategy of eclipse mapping could be refined when a larger number of such eclipsing binaries in various orbits will be modeled. It will be interesting to see how diverse or similar the magnetic activity on the surfaces of giant stars could be, during the fast evolution on the red giant branch of the HRD.

\begin{acknowledgements}
We are grateful to an anonymous referee for the remarks and advices which helped us to make the paper more clear. This work was supported by the Hungarian National Research, Development and Innovation Office grants OTKA KH-130526 and the \'Elvonal grant KKP-143986.
BS would like to thank the organizers of the 5th and 6th Phoebe Workshops for creating a supportive and collaborative environment to learn about binary stars.
AH~and~MT~acknowledge support from the Jenny and Antti Wihuri Foundation.
This work made extensive use of \texttt{numpy} \citep{numpy}, \texttt{scipy} \citep{scipy}, \texttt{pandas} \citep{mckinney-proc-scipy-2010} and \texttt{matplotlib} \citep{matplotlib}.
Some of our results were based on observations obtained with the Samuel Oschin Telescope 48-inch and the 60-inch Telescope at the Palomar
Observatory as part of the Zwicky Transient Facility project. ZTF is supported by the National Science Foundation under Grants No. AST-1440341 and AST-2034437 and a collaboration including current partners Caltech, IPAC, the Oskar Klein Center at
Stockholm University, the University of Maryland, University of California, Berkeley , the University of Wisconsin at Milwaukee, University of Warwick, Ruhr University, Cornell University, Northwestern University and Drexel University. Operations are conducted by COO, IPAC, and UW.  This research has made use of the NASA/IPAC Infrared Science Archive, which is funded by the National Aeronautics and Space Administration and operated by the California Institute of Technology through DOI 10.26131/IRSA537.

\end{acknowledgements}

\bibliography{aa53772-25}

\begin{appendix}

\section{Time-series spot modeling - details}\label{ts_models}

\subsection{Spot latitudes}\label{spotlatitudes}

Experiments were made using free latitudes to check if we could extract some latitude information of spots based on continuous and high precision data. \citet{1997A&A...323..801K} showed that in the case of highly accurate light curves recovering spot latitudes might be possible to some extent.

\begin{figure}[thb]
    \includegraphics[width=\columnwidth]{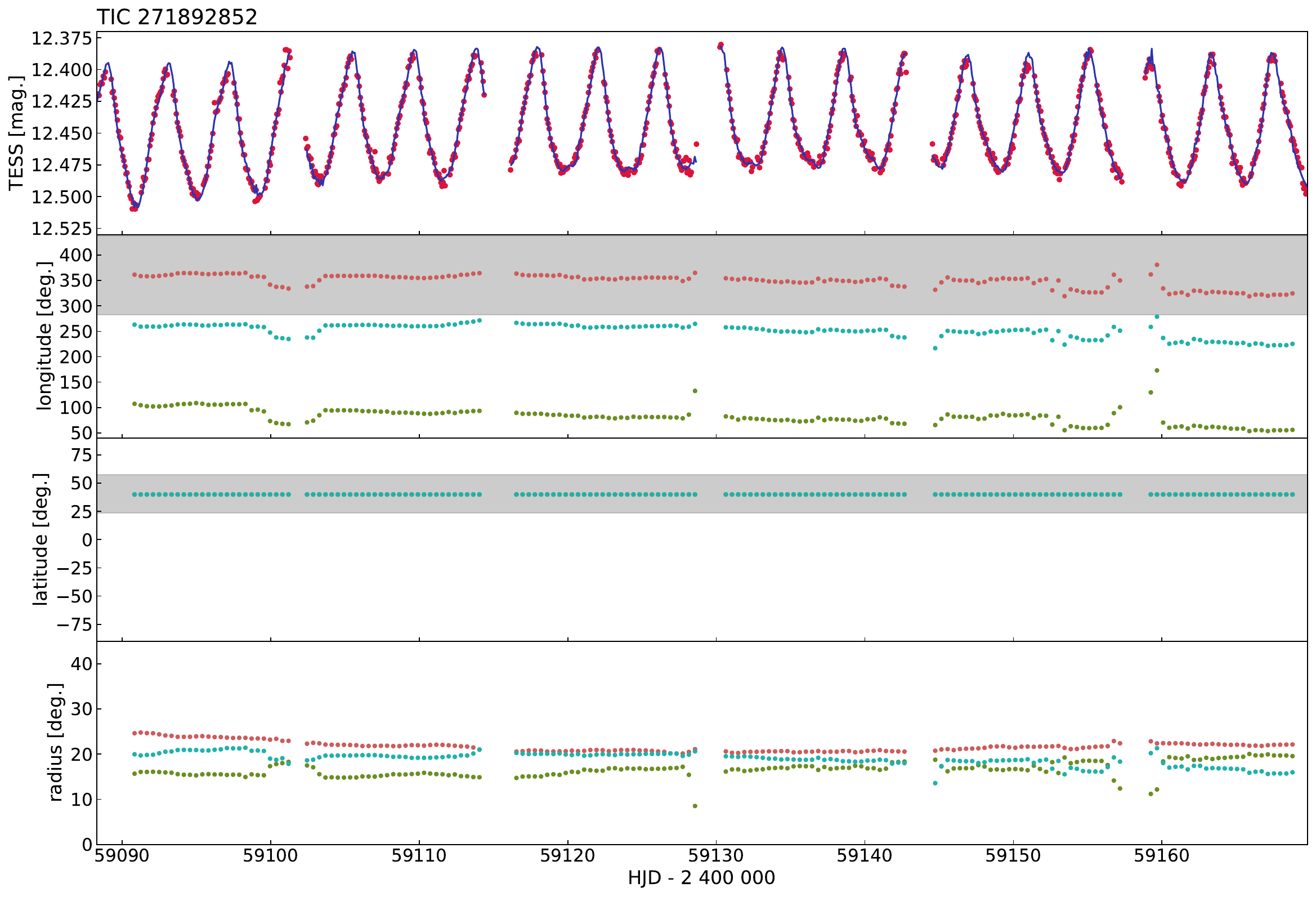}\\
    \includegraphics[width=\columnwidth]{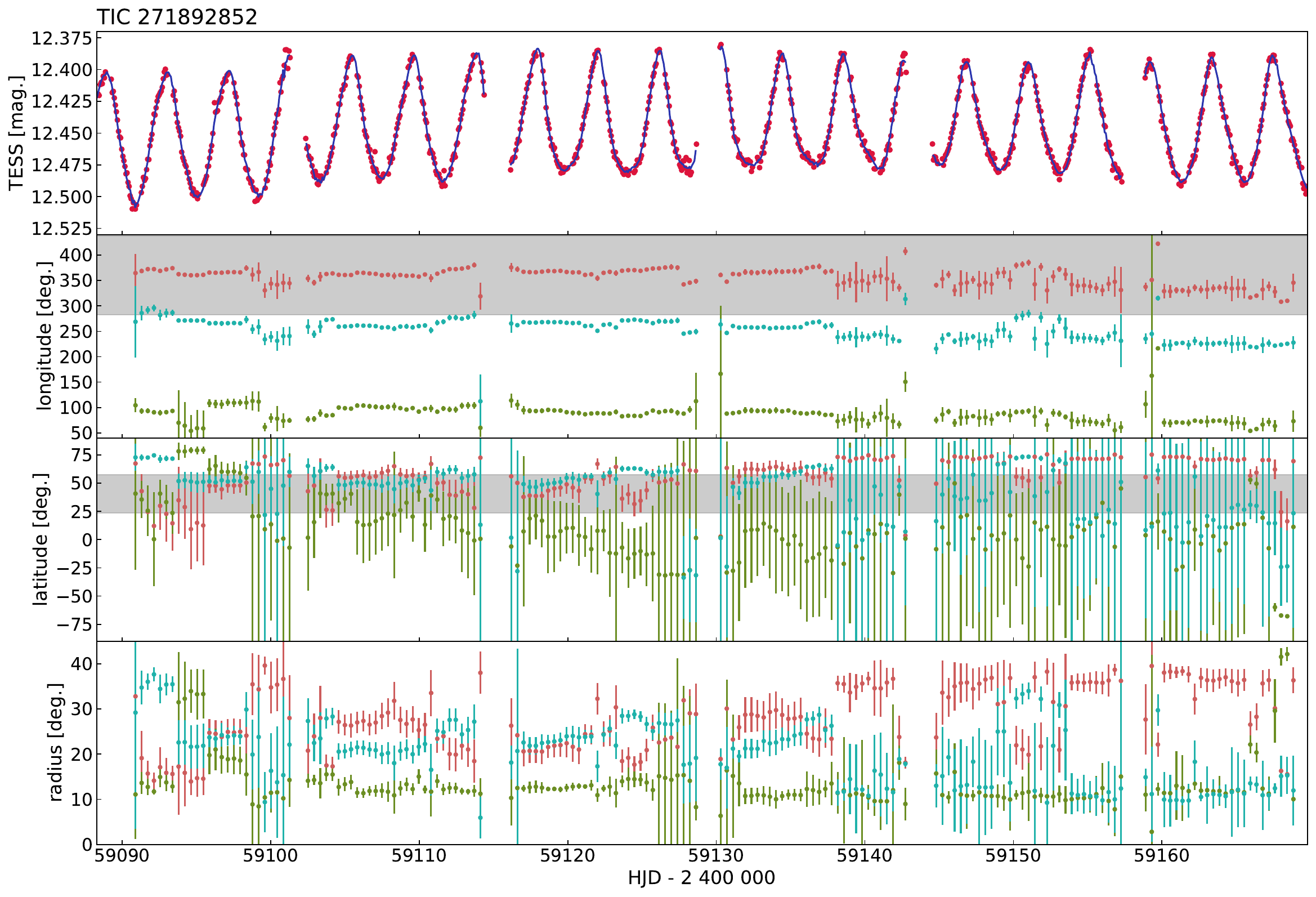}
      \caption{Part of the data of TIC\,271892852 fitted by three spots with fixed (upper panel) and free (lower panel) latitudes. Gray shading marks the scanned longitudes and latitudes by the secondary star. }
      \label{2718_fits}
\end{figure}

\begin{figure}[thb]
    \includegraphics[width=\columnwidth]{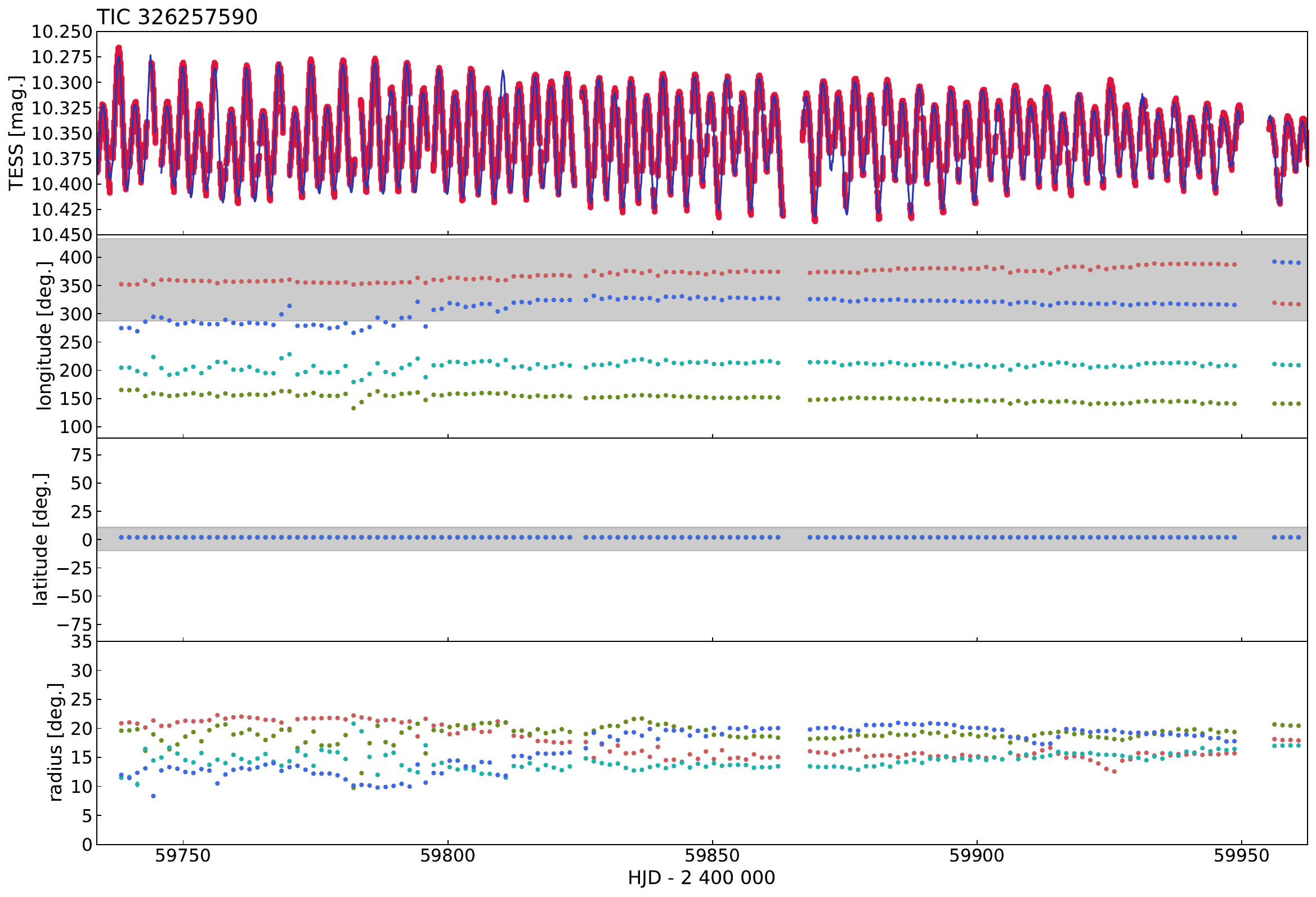}\\
    \includegraphics[width=\columnwidth]{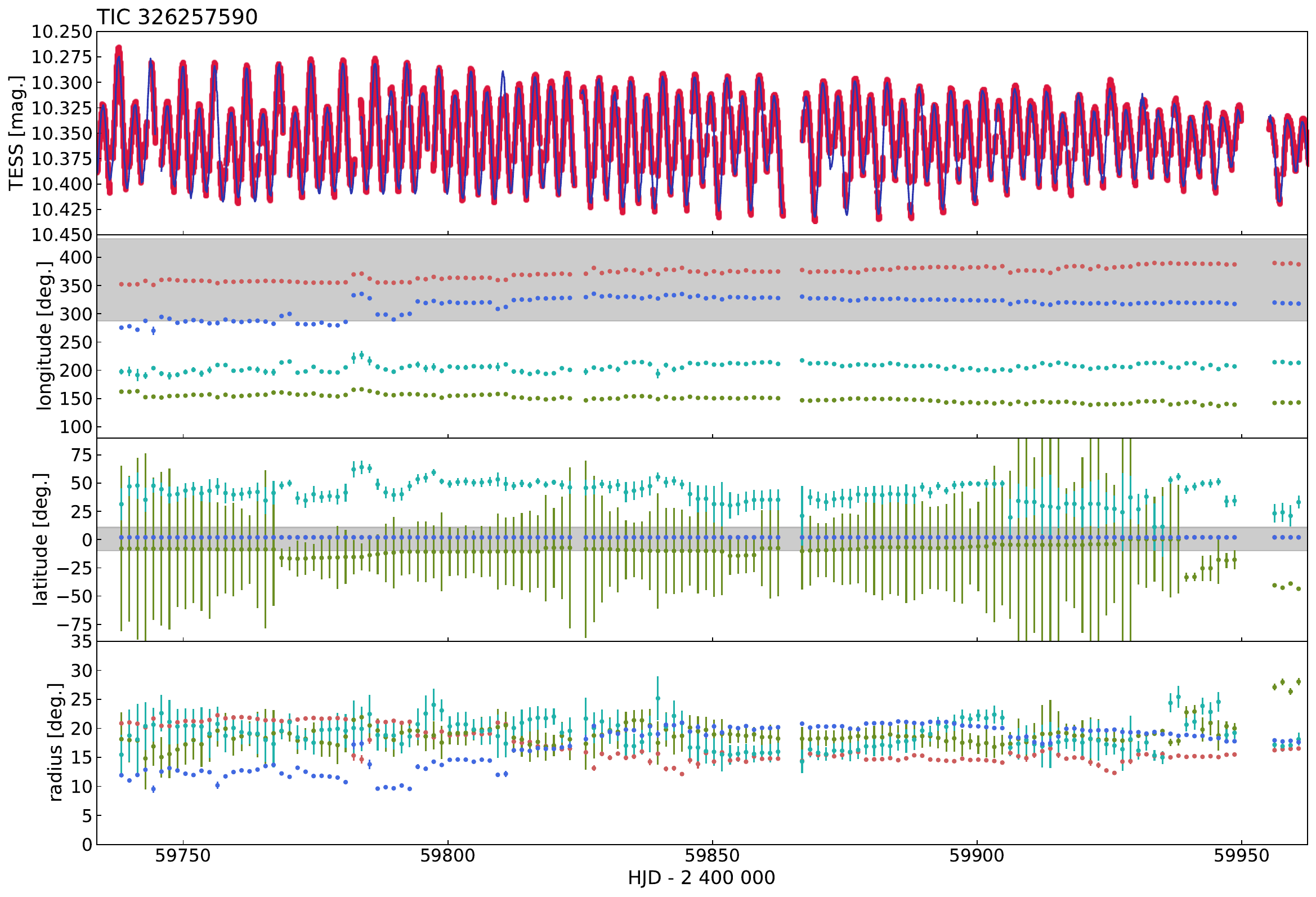}
      \caption{Part of the data of TIC\,326257590 fitted by four spots with fixed (upper panel) and  two free (lower panel) latitudes. Gray shading mark the scanned longitudes and latitudes by the secondary star.}
      \label{3262_fits}
\end{figure}

For the time-series modeling of TIC\,271892852 three spots were used because of the somewhat higher scatter of the data (Fig.\,\ref{2718_lc_examples}). The latitudes were fixed to 40$^{\circ}$ as a mean latitude of the scanned latitudes between $23\fdg7$ and $57\fdg4$ resulting from the eclipse mapping (see Sect.\,\ref{2718_results}). The spot solutions with fixed latitudes (six free parameters, that is, the three longitudes plus the three radii) and spots with free latitudes (additionally three latitudes) are plotted in Fig.\,\ref{2718_fits}. Comparing the two sets of solutions the solid stability of the spot longitudes is evident. As to the latitudes in the lower panel, during the first half of the observations, the spot identified also by the eclipse mapping (marked with red, in the scanned zone) tends to situate in high latitudes, especially where the observational scatter is lower and there is no missing data. Still, the errors of the latitude values are high, from which higher error of spot sizes are also follows. The coordinates of spots at all time steps are independent modeling results with starting latitudes near the equator (+2$^{\circ}$).

TIC\,326257590 is about 2 magnitudes brighter than TIC\,271892852, and TESS provided very accurate data (Fig.\,\ref{3262_lc_examples}). Eclipse mapping usually revealed 2--3 spots near the equator. Fitting the time-series light curves four spots were necessary. In the experiment shown in Fig.\,\ref{3262_fits} we assumed fix latitudes near the equator where the secondary star scans the primary (+2$^{\circ}$, upper panel), and two fixed and two free latitudes (lower panel) when checking the stability of the spot coordinates, which in this case means ten free parameters (four longitudes, four radii and two latitudes). The result in the lower panel of Fig.\,\ref{3262_fits} shows that one spot would be located at higher latitude showing acceptable uncertainty, and the spots sizes also look stable.

\subsection{Number of fitted spots}\label{spotnumber}

For time-series spot modeling we use appropriate segments (windows) which are folded with the orbital/rotational period. The bad data, the eclipses, and the occasional flares are removed from the observations leaving behind accurate light curves but with gaps. The windows are at least one rotation long in time, and can be longer to have good enough phase coverage. However it should not be too long, since the rotational light curve is continuously changing. For the folded subset a spot model is calculated, and after that the window is shifted by some phase of the light curve (it is usually between 0.1$-$0.25 in rotational phase). The process is repeated until the end of the dataset. The windows and shifts are of different lengths of each dataset taking into account the gaps and the data quality. It is supposed that the resulting parameters are valid in the modeled interval (window), therefore its length should be kept as short as the data coverage allows. Finally, from the resulting parameters of each segment an interpolated fit of the whole dataset is calculated and plotted. 

\begin{figure}[thb]
    \includegraphics[width=9cm]{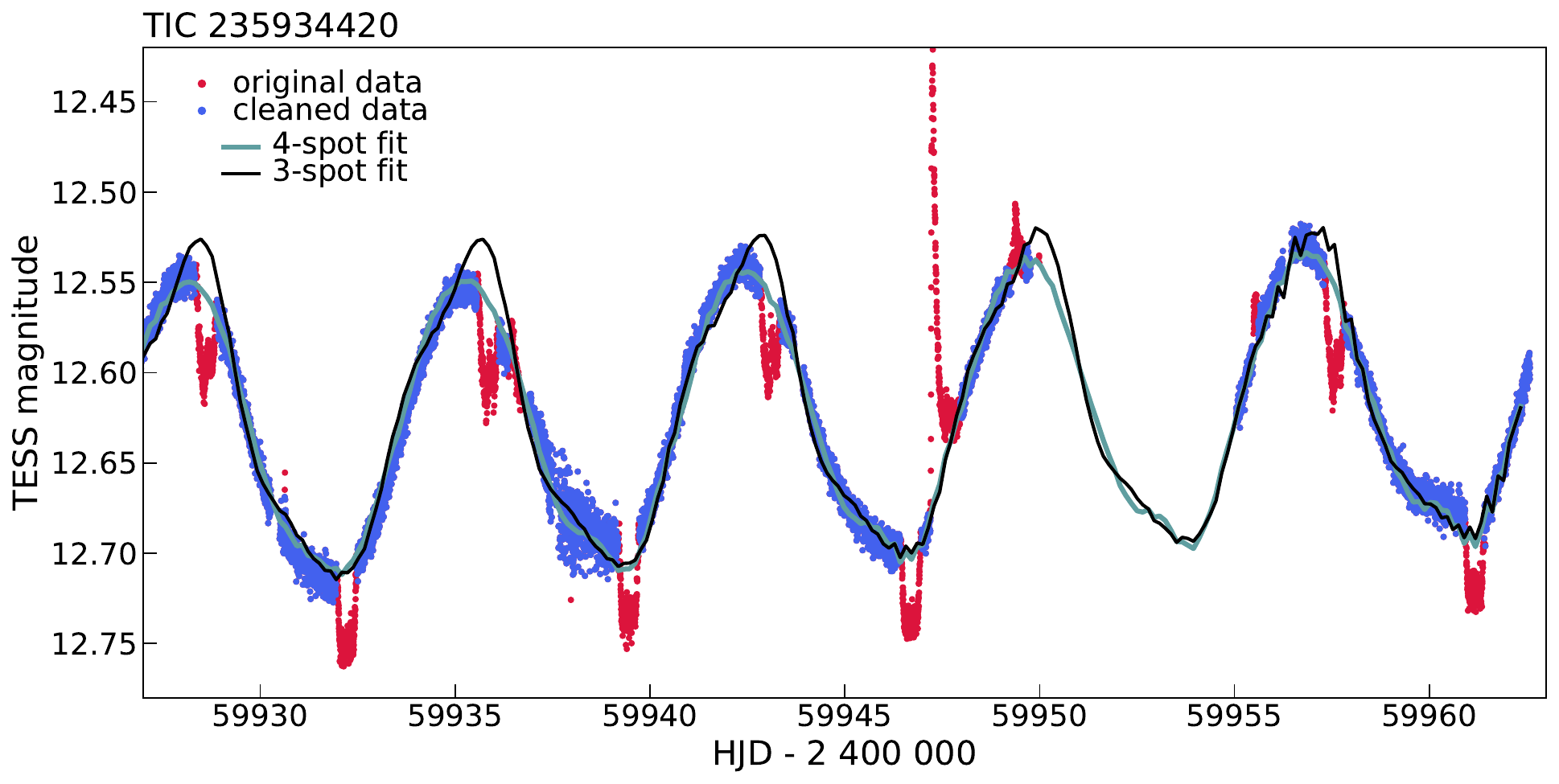}
      \caption{A selected part from the data of TIC\,23934420 showing the difference between three- and four-spot fits.}
      \label{3-4spots}
\end{figure}

To find the suitable window is a delicate problem depending on the data quality and the completeness of the dataset. Choosing unnecessary long data segments may wash out the changes of the spot parameters. In the ideal case of high S/N data without interruption, one rotational cycle is modeled. Figure\,\ref{3-4spots} gives a comparison between the three- and four-spot solutions for a three-rotation long part from a longer set of data of TIC\,23934420. In this case we had to use data from three full rotations as a window, because of the gaps by flares, eclipses and missing data. The step was 0.2 rotational phase. It is well seen that with three spots the light curve could not be fitted well. 

\onecolumn

\section{Two-color modeling of $g$ and $r$ data of TIC\,235934420}\label{ZTF_two-color}

\begin{figure*}[thb]
   \includegraphics[width=18cm]{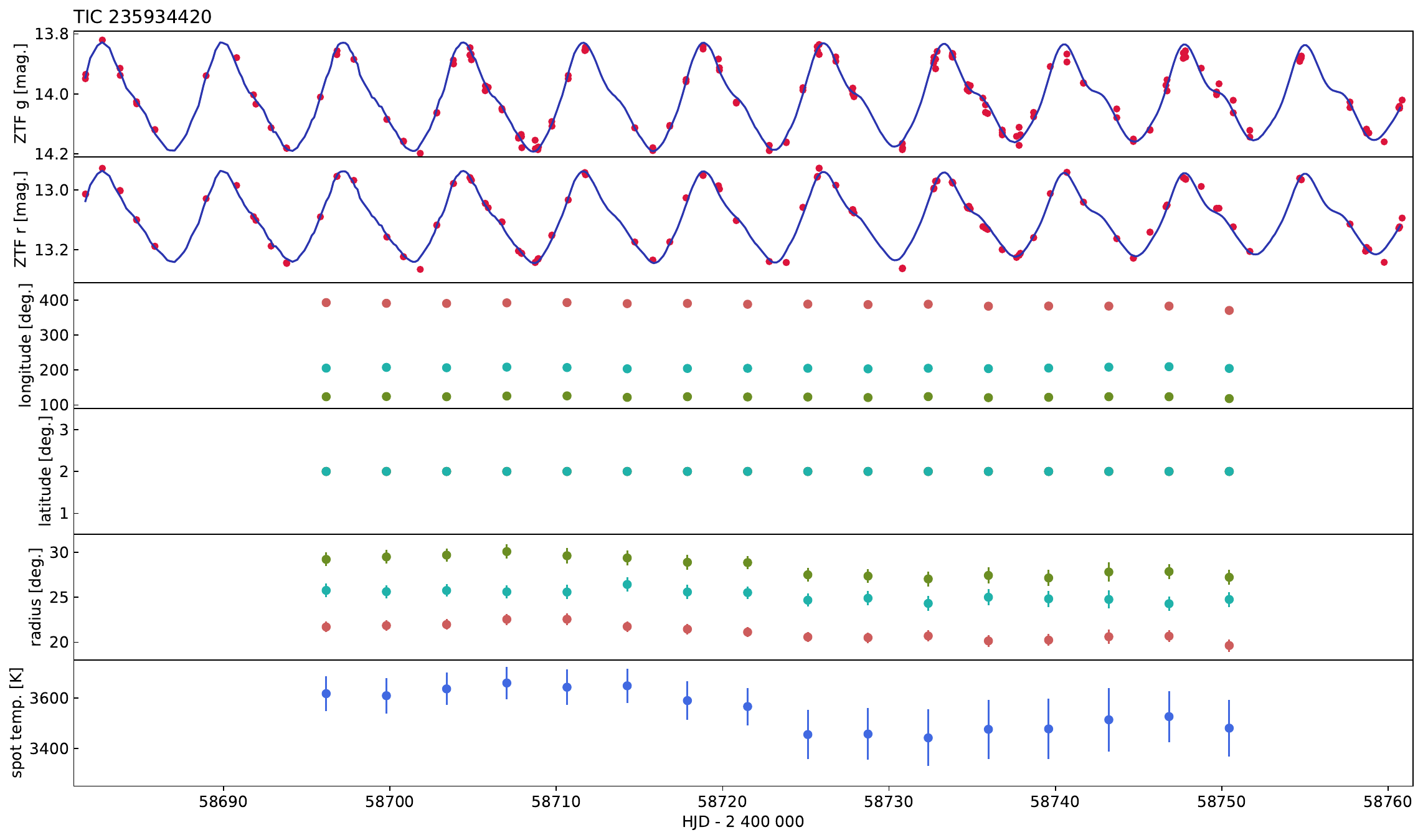}
      \caption{Fitted ZTF $g$ and $r$ data and the resulting spot and temperature parameters. From top to bottom: ZTF $g$ and $r$ observations are shown with their fits. Further below the spots' longitudes, latitudes (fixed) and radii, with their errors, and the resulted spot temperatures and errors, are plotted.}
      \label{2359_ZTF_tser}
\end{figure*}

Taking into account the distance of 886 pc to the system we corrected the magnitudes of TIC\,235934420 with the interstellar extinction. At the distance of the system we find $E(B-V)=0.071$ from NASA/IPAC Infrared Science Archive (IRSA, \mbox{\url{https://irsa.ipac.caltech.edu}}). With the help of the equations in \cite{2012ApJS..199...30P} we get $A_g=0.26$ and $A_r=0.19$ which were used as corrections to the ZTF $g$ and $r$ magnitudes. The direct result from \mbox{\url{http://argonaut.skymaps.info}} is \mbox{$E(g-r)=0.08$}, in good agreement with $E(g-r)=0.07$ from IRSA. The results of the time-series modeling of the two-color ZTF data are plotted in Fig.\,\ref{2359_ZTF_tser} with the standard deviation of the parameters. While the spot longitudes are stable, a systematic change after about HJD\,2458715 is visible in all spot sizes parallel with decreasing spot temperature, indicating that smaller and cooler spots describe the light curve. However, due to the limited information of the photometric data, the instability of the process, and the large error in the resulting temperatures, this feature should be taken with caution. In modeling the data a mean spot temperature of 3550$\pm$80\,K is used.

\newpage

\section{Rotational periods of TIC\,326257590 from DASCH, ASAS~SN and TESS data}\label{32_rotper}

\begin{figure*}[thb]
    \includegraphics[width=18cm]{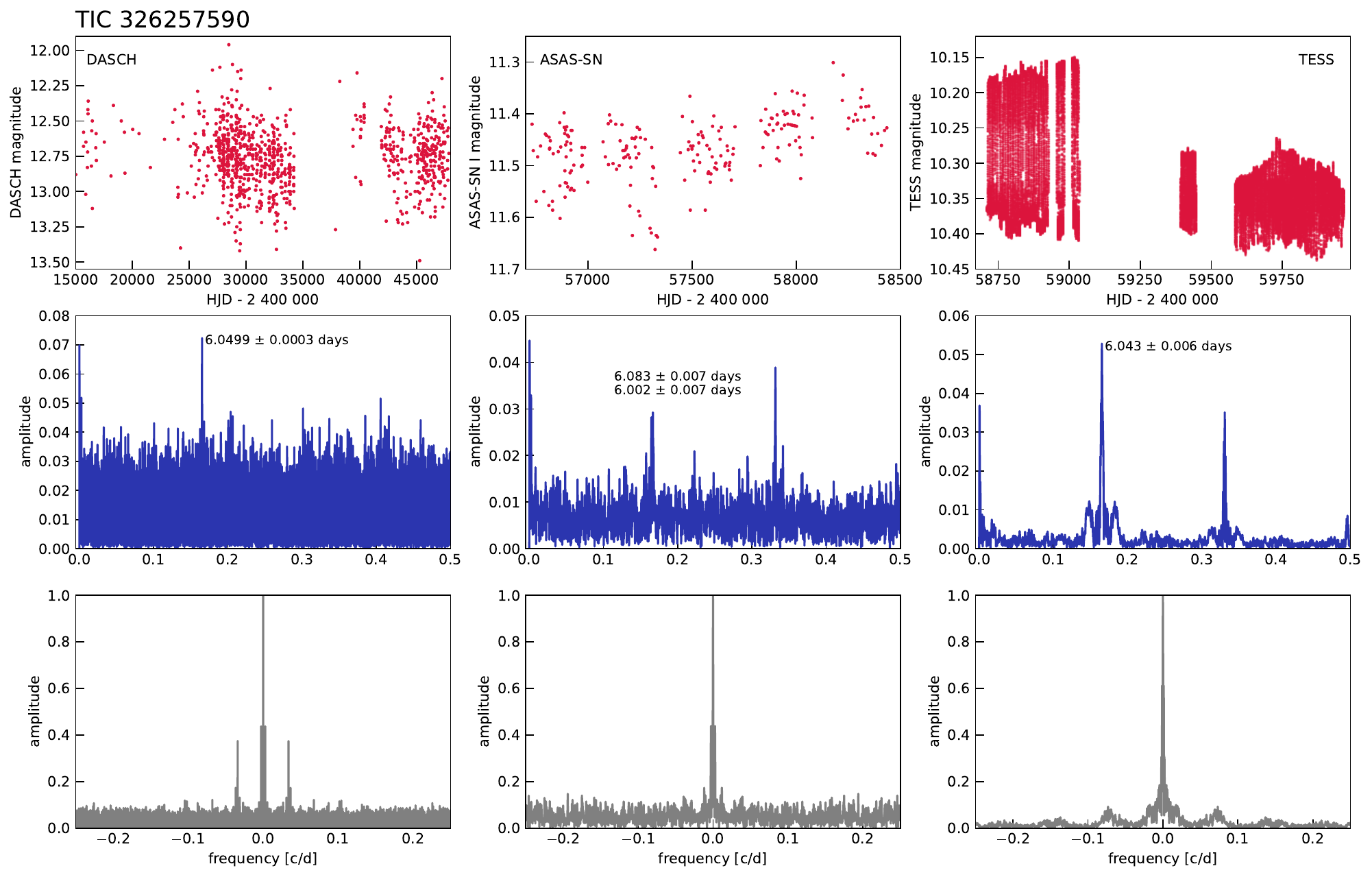}
    \caption{Rotational periods of TIC\,326257590 from DASCH, ASAS-SN and TESS data. In the upper panels the observations are plotted, while in the middle and bottom panels the amplitude spectra and the corresponding spectral windows (gray) are shown, respectively. }
      \label{3262_rotper}
\end{figure*}

\end{appendix}

\end{document}